\documentclass[numberedappendix]{emulateapj}

\usepackage{amsmath}

\shorttitle{Low Virial Parameters in Molecular Clouds}
\shortauthors{Kauffmann, Pillai \& Goldsmith}

\def\tCO {$\mathrm{^{13}CO}$} 

\def\gcmsq  {$\mathrm{g\,cm^{-2}}$}    

\def\AMM {$\mathrm{NH_3}$} 
\def\DAMM {$\mathrm{NH_2D}$} 
\def\NTH {$\mathrm{N_2H^{+}}$} 
\def\NTD {$\mathrm{N_2D^{+}}$} 

\def\simgreat{\mathbin{\lower 3pt\hbox
     {$\rlap{\raise 5pt\hbox{$\char'076$}}\mathchar"7218$}}}
\def\simless{\mathbin{\lower 3pt\hbox
     {$\rlap{\raise 5pt\hbox{$\char'074$}}\mathchar"7218$}}}

\renewcommand{\d}{\ensuremath{\,{\rm{}d}}}

\begin{document}

\title{Low Virial Parameters in Molecular Clouds:\\
  Implications for High Mass Star Formation and
  Magnetic Fields}

\author{Jens Kauffmann\altaffilmark{1}, Thushara Pillai\altaffilmark{1}}

\affil{Astronomy Department, California Institute of Technology, 1200
  East California Blvd., Pasadena, CA 91125, USA}

\altaffiltext{1}{Both authors contributed equally to this work.}

\author{Paul F.\ Goldsmith}

\affil{Jet Propulsion Laboratory, California Institute of Technology,
  4800 Oak Grove Blvd., Pasadena, CA 91109, USA}

\email{tpillai@astro.caltech.edu, jens.kauffmann@astro.caltech.edu}

\begin{abstract}
  Whether or not molecular clouds and embedded cloud fragments
  are stable against collapse is of utmost importance for the study of
  the star formation process. Only ``supercritical'' cloud fragments
  are able to collapse and form stars. The virial parameter
  $\alpha=M_{\rm{}vir}/M$, which compares the virial to the
  actual mass, provides one way to gauge stability against
  collapse. Supercritical cloud fragments are characterized by
  $\alpha{}\lesssim{}2$, as indicated by a comprehensive
    stability analysis considering perturbations in pressure and
    density gradients. Past research has suggested that virial
  parameters $\alpha{}\gtrsim{}2$ prevail in clouds. This would
  suggest that collapse towards star formation is a gradual and
  relatively slow process, and that magnetic fields are not needed to
  explain the observed cloud structure. Here, we review a range of
  very recent observational studies that derive virial parameters
  $\ll{}2$ and compile a catalogue of 1325 virial parameter
  estimates. Low values of $\alpha$ are in particular observed for
  regions of high mass star formation (HMSF). These observations
  may argue for a more rapid and violent evolution during
  collapse. This would enable ``competitive accretion'' in HMSF,
  constrain some models of ``monolithic
  collapse'', and might explain the absence of high--mass starless
  cores. Alternatively, the data could point at the presence of
  significant magnetic fields $\sim{}1~\rm{}mG$ at high gas
  densities. We examine to what extent the derived observational
  properties might be biased by observational or theoretical
  uncertainties. For a wide range of reasonable parameters, our
  conclusions appear to be robust with respect to such biases.
\end{abstract}


\maketitle


\defcitealias{bertoldi1992:pr_conf_cores}{BM}

\section{Introduction}
Whether or not clouds and embedded cloud fragments\footnote{We use the
  word ``fragment'' to denote substructure in clouds, as explained in
  Sec.~\ref{sec:vp-summary}. Our analysis is not concerned with the
  fragmentation processes that might create such substructure.} are
stable against collapse is of utmost importance for the study of
molecular clouds and the star formation process. ``Subcritical'' cloud
fragments are unbound, and may expand and dissolve back into the
diffuse interstellar medium. Conversely, ``supercritical'' fragments
are bound or marginally gravitationally bound, and can undergo
collapse when perturbed. Such cloud fragments can eventually form
stars. The virial parameter, $\alpha\equiv{}5\sigma_v^2R/(GM)$
(\citealt{bertoldi1992:pr_conf_cores}; hereafter
\citetalias{bertoldi1992:pr_conf_cores}) can be used to gauge whether
a cloud fragment is sub-- or supercritical. It is calculated from a
fragment's mass, radius, and one--dimensional velocity dispersion
($M$, $R$, and $\sigma_v$; $G$ is the constant of gravity). For given
environmental conditions, there is a critical virial parameter such
that subcritical clouds are characterized by
$\alpha>\alpha_{\rm{}cr}$, while $\alpha<\alpha_{\rm{}cr}$ holds for
supercritical clouds. For non--magnetized clouds,
$\alpha_{\rm{}cr}\gtrsim{}2$, while strong magnetic fields imply
$\alpha_{\rm{}cr}\ll{}2$ (see Sec.~\ref{sec:vp-summary} and
Appendix~\ref{app:vp-calculations} for both statements).

\citet{larson1981:linewidth_size} presented some of the earliest
observational work on $\alpha$ that used a large sample containing
several clouds. In his Fig.~4, he examines the ratio
$\sigma_v^2R/(GM)=\alpha/5$. He derives the mean and the dispersion
around the average value, which gives a range
$1.1^{+1.6}_{-0.6}$. This implies $\alpha=5^{+8}_{-3}$ in our
framework. Since that study, virial parameters $\alpha\gtrsim{}2$ are
regularly considered to be a general feature of molecular cloud
structure on any spatial scale. For example,
  \citet{heyer2009:scaling-relations} find a mean value $\alpha=1.9$
  for their cloud sample. In fact, the apparent observation that
clouds and cloud fragments are critical or subcritical is commonly
known as ``Larson's second law'' of cloud structure\footnote{We stress
  that \citet{larson1981:linewidth_size} intends to understand
  order--of--magnitude effects of cloud structure. Within the factor
  $\sim{}10$ range considered by him, his and our results are
  consistent.} (e.g., \citealt{mckee2007:review}).

The supposed prevalence of virial parameters $\alpha\gtrsim{}2$ has a
range of consequences for star formation physics. First, since cloud
fragments do not seem to reside in the highly unstable domain
$\alpha{}\ll{}2$, contraction towards collapse is supposedly gradual
and does not occur with free--fall velocities. Second, if high--mass
cores can be modelled as non--magnetized hydrostatic spheres supported
by ``turbulent'' gas motions with velocity dispersion $\sigma_v$, then
stellar accretion rates $\dot{M}_{\star}\propto{}\sigma_v^3$ during
collapse are implied (e.g.,
  \citealt{shu1977:self-sim_collapse}). Given the large observed
values of $\sigma_v$, this would offer a straightforward explanation
why high--mass stars can continue to accrete despite exerting
significant radiation pressure on their environment
\citep{mckee2002:massive-sf}. Third, dynamically significant magnetic
fields would not be needed to explain the structure of molecular
clouds. If $\alpha{}\gtrsim{}2$, gas motions alone could prevent cloud
fragments from collapsing, while $\alpha{}<2$ would imply a
significant role for magnetic fields in star formation (e.g.,
\citealt{myers1988:mhd-equilibrium}). Fourth, star formation via
``competitive accretion'' would not work, since this requires high
densities and slow relative motions between stars and gas (implying
$\alpha<1$; \citealt{krumholz2005:non-competitive}).

We remark that the virial parameter is also important when one wishes
to understand the evolution of entire molecular clouds. Virial
parameters in large cloud complexes were studied by, e.g.,
\citet{solomon1987:scaling-relations},
\citet{scoville1987:inner-galaxy-clouds}, and
\citet{heyer2001:equilibrium}. Recent updates are provided by
\citet{heyer2009:scaling-relations} and
\citet{roman-duval2010:grs-clouds}. 
\citet{dobbs2011:unbound-clouds} suggest that clouds in their entirety
are unbound on large spatial scales $\gtrsim{}10~\rm{}pc$. This would
present a convenient explanation for the low star formation rate in the
Milky Way.\medskip

For these reasons, it is important to realize that a number of
observational studies conducted in the last few years \emph{do} find
massive cloud fragments with $\alpha{}\ll{}2$ (e.g.,
\citealt{pillai2011:initial-conditions},
\citealt{csengeri2011:cygnus-x}, \citealt{wienen2012:ammonia},
\citealt{li2012:orion}). These observations also mean that constraints
on star formation physics based on the assumption of
$\alpha{}\gtrsim{}2$ have to be reconsidered. Returning to the list
above, collapse might be rapid and violent. There is little evidence
to justify spheres in hydrostatic equilibrium supported by turbulent
pressure, for which $\dot{M}_{\star}\propto{}\sigma_v^3$ would
hold. Competitive accretion may occur, and significant magnetic fields
might be needed to explain cloud structure.

More generally, the new observations constrain under which conditions
numerical cloud simulations with initial virial parameters $\alpha>1$
(e.g., series of simulations started by
\citealt{bate2002:brown-dwarfs}) or $\alpha<1$ (e.g.,
\citealt{hennebelle2011:collapse-outflows-fragmentation}) may apply,
and why some stellar clusters might be born with velocity dispersions
too low to balance self--gravity (see \citealt{adams2010:review} for a
summary). The compilation presented here also gauges to what extent
virial masses can be used to estimate true fragment masses.

However, before discussing the implications of small virial
parameters, it is prudent to re--examine observational determinations
of $\alpha$. Here, we do so by presenting a large catalogue of virial
parameter estimates generated by reprocessing a wide variety of
previously published observations in a standardized fashion. This
helps to separate true observational trends from artifacts found in
smaller samples. Also, past studies calculated virial parameters using
a very broad range of definitions for $\alpha$. This means that virial
parameters reported by different studies can usually not be directly
compared with another.

Further, it is important to be aware of the model assumptions
determining the value of $\alpha_{\rm{}cr}$, which controls whether an
observed cloud fragment is stable or
not. \citet{ballesteros-paredes2006:six-myths} explores some of the
most common misconceptions about the virial parameter. In particular,
he stresses that there is constant mass flow between all scales, so
that a static picture of cloud structure may not be appropriate. That
said, he concludes that ``the sub-- or supercriticality of a molecular
cloud core [judged based on the virial mass ratio] is a good
estimation of the dynamical state of such a core'', in accordance with
the assumptions made in the current paper.\medskip

We address these points as follows. First, Sec.~\ref{sec:vp-summary}
summarizes the concept of the virial parameter, and reviews the
expected critical values controlling the stability of cloud fragments
against collapse. Section~\ref{sec:analysis} presents the reprocessing
of existing observational data to derive a catalogue of virial
parameter estimates. Trends found within this catalogue are described
and discussed in Sec.~\ref{sec:trends}. In particular, this shows that
$\alpha{}\ll{}2$ is found in a variety of
samples. Section~\ref{sec:sf-physics} describes possible implications
for star formation physics. Whether observational uncertainties could
bias the virial parameter to values $\ll{}2$ is examined in
Sec.~\ref{sec:vp_correct-relevant}. The paper concludes with a summary
in Sec.~\ref{sec:summary}. In three appendices, we examine the virial
parameter (Appendix~\ref{app:vp-calculations}), its dependency on the
fragment masses (Appendix~\ref{app:vp-slope}), and implications from
low observed virial parameters (Appendix~\ref{app:implications}), in
more detail.

Sections \ref{sec:samlple-analysis} and \ref{sec:trends} are of a
somewhat technical nature. These parts of the discussion may be
skipped if one wishes to proceed directly to the essential parts of
the paper.

\section{The Virial Parameter: An Overview}
For more than two decades, the virial parameter and the related virial
mass have been employed to gauge whether or not a cloud fragment is
stable against collapse. Here, we define a few relevant properties and
summarize how the virial parameter can be used to gauge whether a
cloud fragment is unstable to gravitational collapse or not. Some of
the details of the discussion are deferred to
Appendix~\ref{app:vp-calculations}.\label{sec:vp-summary}

Throughout this paper, we consider ``cloud fragments'' as entities of
arbitrary size that form part of larger molecular clouds. Fragments
include, for example, the ``dense cores'' and ``clumps'' discussed in
other studies (e.g., \citealt{williams2000:pp_iv}).\medskip

We execute our analysis in the framework laid out by
\citeauthor{bertoldi1992:pr_conf_cores}
(\citeyear{bertoldi1992:pr_conf_cores}; hereafter
\citetalias{bertoldi1992:pr_conf_cores}). They define the virial
parameter as
\begin{subequations}
\label{eq:virial-parameter}
\begin{align}
  \alpha & \equiv \frac{5\sigma_v^2R}{GM}
  \label{eq:virial-parameter-definition} \\
  & = 1.2\,\left(\frac{\sigma_v}{\rm{}km\,s^{-1}}\right)^2
  \left(\frac{R}{\rm{}pc}\right)
  \left(\frac{M}{10^3\,\rm{}M_{\sun}}\right)^{-1}\,{}.
\end{align}
\end{subequations}
One may simplify this and write $\alpha=M_{\rm{}vir}/M$ by defining a virial mass
$M_{\rm{}vir}\equiv5\sigma_v^2R/G$.  As detailed in
Sec.~\ref{sec:analysis}, the velocity dispersion combines the thermal
motion of the mean free particle in interstellar molecular gas with
the non--thermal motions of the bulk gas. This definition of $\alpha$ was chosen
  since---as shown by \citetalias{bertoldi1992:pr_conf_cores}---
\begin{equation}
\alpha=a\,\frac{2E_{\rm{}kin}}{\left|E_{\rm{}pot}\right|}\,{},
\label{eq:energy-ratio}
\end{equation}
where $E_{\rm{}kin}$ and $E_{\rm{}pot}$ are the kinetic and
gravitational potential energy, respectively. The parameter $a$
absorbs modifications that apply in the case of non--homogeneous and
non--spherical density distributions. Evaluation gives $a=2\pm{}1$
for a wide range of cloud shapes and density gradients (see
Appendix~\ref{app:vp-calculations}).

The fact that $\alpha$ is related to
$E_{\rm{}kin}/\left|E_{\rm{}pot}\right|$ can be used for a \emph{naive
  stability analysis that neglects a few details}. Cloud fragments
with $E_{\rm{}kin}/\left|E_{\rm{}pot}\right|\gg{}1$, and therefore
$\alpha{}\gg{}1$, contain enough kinetic energy to expand and dissolve
into the surrounding environment. Alternatively, they may also be
confined by additional forces, such as an external pressure (see,
e.g., the ``pressure confined'' fragments discussed by
\citetalias{bertoldi1992:pr_conf_cores}). Conversely, fragments with
$E_{\rm{}kin}/\left|E_{\rm{}pot}\right|\ll{}1$, and therefore
$\alpha{}\ll{}1$, do not hold significant kinetic energy, and will
often not be stable: they will collapse or must be supported against
self--gravity. This suggests the existence of a critical virial
parameter $\alpha_{\rm{}cr}\sim{}1$. ``Supercritical'' fragments with
$\alpha<\alpha_{\rm{}cr}$ will collapse, while ``subcritical''
fragments with $\alpha>\alpha_{\rm{}cr}$ will expand or must be
confined.

More detailed models of the stability of cloud fragments usually
consider a models' response to perturbations. For example, isothermal
hydrostatic equilibrium spheres, as discussed by
\citet{ebert1955:be-spheres} and \citet{bonnor1956:be-spheres}, are
stable against slow perturbations provided their mass is below the
critical value, $M<M_{\rm{}BE}$, where
\begin{equation}
M_{\rm{}BE} = 2.43\,\frac{\sigma_v^2\,R}{G}\,.
\label{eq:be-mass}
\end{equation}
Substitution of this mass into Eq.~(\ref{eq:virial-parameter}) gives
$\alpha_{\rm{}BE}=2.06$, the critical virial parameter of
Bonnor--Ebert spheres. As laid out in
Appendix~\ref{app:vp-calculations}, $M_{\rm{}BE}$ actually provides an
approximate upper limit to the critical masses of non--magnetized
spheres, provided that the pressure comes from random motions with a
velocity dispersion increasing outwards (as observed; see
Sec.~\ref{sec:trends}): $M_{B=0}\lesssim{}M_{\rm{}BE}$. It follows
that
\begin{equation}
\alpha_{B=0} \gtrsim \alpha_{\rm{}BE} \approx 2
\end{equation}
holds for the critical virial parameter of a hydrostatic equilibrium
sphere not supported by magnetic pressure. For reference, the velocity
dispersion needed to achieve $\alpha=2$ is presented in
Sec.~\ref{sec:sf-physics}.

For magnetized clouds, a critical mass $M_B\approx{}M_{B=0}+M_{\Phi}$
holds \citepalias{bertoldi1992:pr_conf_cores}, where the magnetic flux
mass for a field of mean strength $\langle{}B\rangle$ is
\begin{equation}
M_{\Phi} = 0.12\,\frac{\Phi}{G^{1/2}} =
   0.12\,\frac{\pi{}\langle{}B\rangle{}R^2}{G^{1/2}}
\label{eq:flux-mass}
\end{equation}
\citep{tin_ii}. $M_{B=0}\lesssim{}M_{\rm{}BE}$ implies that
$M_B\lesssim{}M_{\rm{}BE}+M_{\Phi}$. One may further rewrite
Eq.~(\ref{eq:virial-parameter-definition}), without loss of
generality, as $\alpha=\alpha_{\rm{}BE}\cdot{}(M_{\rm{}BE}/M)$. In the
critical case, with $M=M_B$, substitution of
$M_B\lesssim{}M_{\rm{}BE}+M_{\Phi}$ and rearrangement gives
\begin{equation}
\alpha_B \gtrsim 2 \cdot
  \frac{1}{1+M_{\Phi}/M_{\rm{}BE}}\,,
\label{eq:alpha_cr-magnetic-approx}
\end{equation}
where we have further used that $\alpha_{\rm{}BE}\approx{}2$. In the limit
$M_{\Phi}\gg{}M_{\rm{}BE}$, one finds
$\alpha_B=\alpha_{\rm{}BE}\cdot{}(M_{\rm{}BE}/M_{\Phi})$. This shows that
$\alpha_B$ can attain values $\ll{}2$, provided the magnetic field is
strong enough.\medskip

The discussion above assumes roughly spherical models with moderate
density gradients. In principle, deviations from spherical shape and
variations in density gradients will affect a model's stability
against perturbations. However, since $a$ is approximately constant
over a wide range of parameters (Appendix~\ref{app:vp-calculations}),
moderate deviations from the assumed density profiles should not
significantly affect the energy ratios, and therefore also not affect
the stability considerations.

We note that the impact of inhomogeneous density distributions
\emph{does not have to be considered} when comparing observed virial
parameters against specific model values, such as
$\alpha_{\rm{}BE}$. \emph{Detailed model calculations, e.g.\ of the
  Bonnor--Ebert case, already take density gradients into account.}
Correction of the observed $\alpha$, or of the model value
$\alpha_{\rm{}cr}$, would thus be overcorrecting for density
gradients.

\begin{figure*}
\centerline{\includegraphics[width=0.6\linewidth]{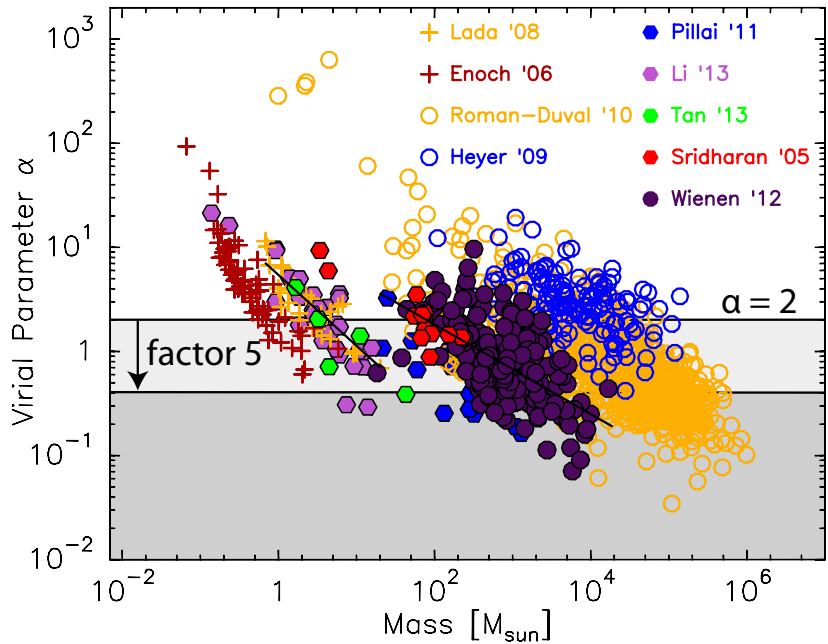}}
\caption{The virial parameter, $\alpha$, as a function of fragment
  mass. Filled symbols indicate samples focusing on regions of high
  mass star formation (HMSF), while crosses indicate studies dealing
  with non--HMSF regions. Open circles are used for mixed samples that
  include clouds with and without HMSF. The horizontal black line at
  $\alpha=2$ gives the lowest critical virial parameter expected for
  non--magnetized clouds: fragments with $\alpha<2$ are supercritical
  and unstable to collapse, unless they are supported by significant
  magnetic fields. For reference, a virial parameter lower by a factor
  5 is indicated using an arrow and another horizontal line. To
  illustrate power--law trends in virial parameters, fits to the
  \citet{lada2008:pipe-nature} and \citet{wienen2012:ammonia} samples
  are drawn using black lines.\label{fig:vp-observed}}
\end{figure*}

\section{Sample Selection \& Data Analysis}
This section explains how $\alpha$ is calculated from the data. The
results are shown in Fig.~\ref{fig:vp-observed}. We compile a
catalogue containing a total of 1325 virial parameter estimates for
entire molecular clouds ($\gg{}1~\rm{}pc$ scale), clumps
($\sim{}1~\rm{}pc$), and cores ($\ll{}1~\rm{}pc$) with and without
high mass star formation (HMSF).\label{sec:samlple-analysis}

As we detail below, observations of entire clouds are taken from
\citet{heyer2009:scaling-relations} and
\citet{roman-duval2010:grs-clouds}. Data for HMSF regions are from
\citet{sridharan2005:high-mass-starless-cores},
\citet{pillai2011:initial-conditions}, \citet{wienen2012:ammonia},
\citet{li2012:orion}, and \citet{tan2013:hmsc}. These samples all
focus on, or exclusively study, early stages of HMSF. This means no or
only faint sources are embedded in the fragments. We choose to
concentrate on such young sources, because the state of more evolved
objects is not necessarily representative of the initial conditions
for star formation. For this reason, we do not include data from HMSF
samples where---as discussed by the respective authors---evolutionary
indicators like masers, mid--infrared emission, outflows, infall line
profiles, etc., imply active star formation
\citep{plume1997:water-maser-cores, molinari2000:hii-precursors,
  beuther2002:hmpo-densities, beltran2004:rotating-disks,
  fontani2005:southern-hmpo-candidates, bontemps2010:cygnus-x,
  csengeri2011:cygnus-x}.  Data for non--HMSF cores include cores in
the Pipe Nebula \citep{lada2008:pipe-nature} and Perseus molecular
cloud \citep{enoch2006:perseus}. Notes on individual samples are given
in Sec.~\ref{sec:sample-details}. Table~\ref{tab:data-summary}
presents an overview over the different studies used here.

\emph{Our combined sample is not meant to be complete and unbiased};
while our guiding principle is to provide a comprehensive overview,
when possible we focus on larger samples that can easily be processed
in the standardized way outlined below. A key aspect is that we
\emph{only} employ masses derived from observations of dust emission
and extinction, and use a common set of dust opacities to derive
masses from these data. This approach is chosen since mass
measurements based on molecular line emission suffer from
uncertainties due to unknown molecular abundances and excitation
conditions. We deviate from this strategy only when considering very
large clouds, for which only observations of molecular line emission
from CO are available. Definitions of radius and velocity dispersion
are standardized for all data, as explained below.\label{sec:analysis}

\subsection{Data Processing}
Several properties must be calculated to estimate the virial
parameter. This is done as follows.

\begin{figure}
\includegraphics[width=\linewidth]{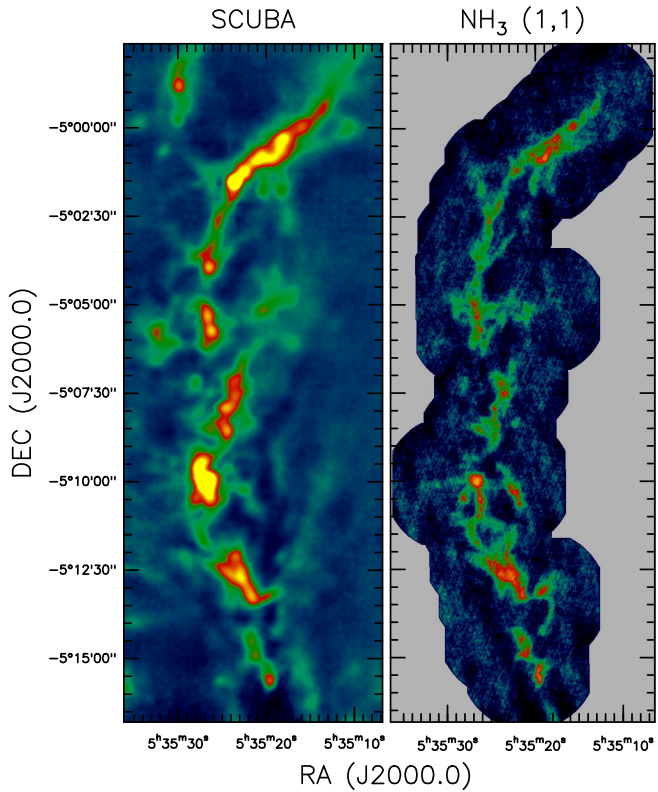}
\caption{Comparison of emission from dust (\emph{left panel}; SCUBA
  data from \citealt{nutter2007:orion}) and Ammonia (\emph{middle};
  combined VLA and GBT data from \citealt{li2012:orion}). The
  correlation between the tracers is obvious. This justifies the use
  of tracers like Ammonia to characterize the kinematics of material
  detected in dust emission.\label{fig:overview_region}}
\end{figure}

\paragraph{Velocity Dispersion, $\sigma_v$} Some studies provide the
FWHM line width (\citealt{roman-duval2010:grs-clouds},
\citealt{wienen2012:ammonia}, \citealt{bontemps2010:cygnus-x},
\citealt{pillai2011:initial-conditions},
\citealt{sridharan2005:high-mass-starless-cores}). In those cases, we
calculate the corresponding velocity dispersion for a Gaussian line
shape. We have discarded data where more than one velocity component
is reported. This avoids the arbitrary choice of how the mass has to
be divided up between the different velocity components.

The velocity dispersion $\sigma_v$ entering the calculation of the
virial parameter reflects the combination of non--thermal gas motions,
$\sigma_{v,\rm{}nt}$, and of the thermal motions of the particle of
mean mass. The latter mass is $\langle{}m\rangle{}=2.33\,m_{\rm{}H}$
for molecular gas at the typical interstellar abundance of H, He, and
metals (Appendix~A of \citealt{kauffmann2008:mambo-spitzer}). For a
molecule of mass $m$, the thermal velocity dispersion at temperature
$T$ is
$\sigma_{{\rm{}th},m}=288~{\rm{}m\,s^{-1}}\cdot{}(m/m_{\rm{}H})^{-1/2}\cdot{}(T/{10~\rm{}K})^{1/2}$,
where $m_{\rm{}H}$ is the hydrogen mass. We combine these relations to
estimate the dynamically relevant velocity dispersion as
$\sigma^2_v=\sigma^2_{{\rm{}th},\langle{}m\rangle}+\sigma^2_{v,\rm{}nt}$. Similarly,
the velocity dispersion observed for a molecule of mass $m$ is a
combination of thermal and non--thermal gas motions,
$\sigma^2_{v,\rm{}obs}=\sigma^2_{{\rm{}th},m}+\sigma^2_{v,\rm{}nt}$. We
use the latter relation to estimate $\sigma_{v,\rm{}nt}$, which is
then used to derive $\sigma_v$.

We have consistently used molecular emission lines selectively tracing
dense gas to estimate velocity dispersions. This assures that the mass
derived from dust emission, and the velocity dispersion from emission
lines, probe the same volume. The dense gas tracers include \AMM,
\DAMM, \NTH, and \NTD, and are not affected by
depletion. Figure~\ref{fig:overview_region} illustrates the strong
correlation between these tracers and the dust emission relevant for
mass measurements. Tracers of lower density gas are only used for the
two $\rm{}^{13}CO$--based cloud samples for which dust continuum data are not
available \citep{heyer2009:scaling-relations,
  roman-duval2010:grs-clouds}. In those cases, we have used
$\rm{}^{13}CO$--derived masses and we have correspondingly derived $\alpha$ using
$\rm{}^{13}CO$ velocity dispersions.

\paragraph{Radius, $R$} We adopt a common definition for the radius
across all samples. Specifically, we determine the source area, $A$,
and convert this into an effective radius, $R=(A/\pi)^{1/2}$. Some
samples report an area contained in a specific contour, or they use
other means to draw a clear outer source boundary. Other publications
report the full width at half maximum (FWHM), and we use the FWHM area
to determine $R$. As described below, we assure that mass and radius
are consistent and refer to the same area.

\paragraph{Mass from Dust Emission, $M$} Most masses are estimated
from dust emission. These masses are calculated from the observed flux
following the formalism from \citet{kauffmann2008:mambo-spitzer},
\begin{equation}
  \begin{array}{l}
    M = 
    \displaystyle 0.12 \, M_{\odot}
    \left( {\rm e}^{1.439 (\lambda / {\rm mm})^{-1}
        (T / {\rm 10 ~ K})^{-1}} - 1 \right) \\
    \quad \displaystyle
    \cdot \left( \frac{\kappa_{\nu}}{0.01 \rm ~ cm^2 ~ g^{-1}} \right)^{-1}
    \left( \frac{F_{\nu}}{\rm Jy} \right)
    \left( \frac{d}{\rm 100 ~ pc} \right)^2
    \left( \frac{\lambda}{\rm mm} \right)^{3}
  \end{array}
\end{equation}
where $\kappa_{\nu}$ is the dust opacity, and $F_{\nu}$ is the integrated
flux for an object at distance $d$ and dust temperature $T$. We adopt
a common opacity law $\kappa_{\nu}$ for dust grains with thin
ice mantles coagulating for $10^5~\rm{}yr$ at density of
$10^6~\rm{}cm^{-3}$ from \citet{ossenkopf1994:opacities}. We adopt a
gas--to--dust ratio of 100. We have used a
power--law slope of 1.75 \citep{battersby2011:cluster-precursors} when
interpolating and extrapolating (for wavelengths
$\lambda>1.3~\rm{}mm$) between tabulated values.

As mentioned before, we assure that mass and radius are consistent
and refer to the same area. Most publications report the flux for the
aforementioned area $A$ used to derive the effective radius. A
different approach is taken by \citet{wienen2012:ammonia} and
\citet{sridharan2005:high-mass-starless-cores}: these authors report
FWHM sizes (\citeauthor{sridharan2005:high-mass-starless-cores} take
those data from \citealt{beuther2002:hmpo-densities}), while the
reported fluxes are integrated over a full Gaussian model fitted to
sources. The latter model exceeds the spatial extent of the FWHM
area. To derive mass and size estimates for the same area, the
mass---respectively the observed flux---is reduced by a factor 0.69
(see \citealt{jenthu2010:irdcs}).

\paragraph{Mass from Dust Extinction, $M$} Core masses in the Pipe
nebula are based on extinction data of \citet{alves2007:imf}. They use
the NICER method on the data from 2MASS bands ($1.25~\mu\rm{}m$,
$1.65~\mu\rm{}m$, $2.2~\mu\mbox{m}$). Core masses are obtained using
an extinction law $A_K/A_V=0.112$ and a conversion factor from
extinction to column density of
$N({\rm{}H_2})/A_V=9.4\times{}10^{20}~\rm{}cm^{-2}\,mag^{-1}$. As
shown in \citet{kauffmann2010:mass-size-i}, based on work by
\citet{bianchi2003:dust-emissivity}, such extinction--based masses are
within a factor $\sim{}1.5$ consistent with masses derived from dust
emission using the aforementioned formalism.

\paragraph{Mass from $^{13}$CO Emission, $M$} The $\rm{}^{13}CO$--based masses are taken
directly from the original publications. These latter mass
measurements are not standardized with respect to the dust
observations and they suffer from other and larger uncertainties
(e.g., in abundance). Also, the interpretation of virial parameters
depends on how exactly the observed $\rm{}^{13}CO$ luminosity traces a cloud's
mass reservoir (e.g., \citealt{maloney1990:virial-equilibrium}). Here,
$\rm{}^{13}CO$--based masses are shown for completeness; they are not used in our
main analysis.

\paragraph{Virial Parameter, $\alpha$} From the properties derived
above, the virial parameter $\alpha$ is eventually calculated in the
same way for all cloud fragments using
Eq.~(\ref{eq:virial-parameter}).

\paragraph{Observational Uncertainties} The aforementioned
observational properties suffer from a variety of observational
uncertainties. Because of the flow of the argument, it is most useful
to discuss these uncertainties in the context of the physical
interpretation of the results. This is done at the end of this
document in Sec.~\ref{sec:vp_correct-relevant}. We briefly note that
the expected mass uncertainties are of order of a factor 2, and are
the largest uncertainty in calculating $\alpha$. The resulting virial
parameters are uncertain by a similar factor.

\subsection{Individual Samples\label{sec:sample-details}}
\paragraph{Giant Molecular Clouds} We have used data from the
$\rm{}^{13}CO$--based survey of giant molecular clouds (GMCs) by
\citet{roman-duval2010:grs-clouds}. They derive masses, sizes and
kinematics for 750 molecular clouds based on the Boston
University--Five College Radio Astronomy Observatory (BU--FCRAO)
Galactic Ring Survey (GRS). This data has overlap with the GRS data
published for 162 GMCs of \citet{heyer2009:scaling-relations}. We also
plot the latter sample for consistency.\footnote{The properties
  derived for the same molecular clouds in both studies (as noted by
  \citeauthor{roman-duval2010:grs-clouds}) show differences because of
  the different methods of structure identification and choice of
  noise threshold. \citeauthor{roman-duval2010:grs-clouds} compute an
  effective radius within the contour detected at the
  $4\sigma$--level, while \citeauthor{heyer2009:scaling-relations} use
  the position centroid and angular extent defined by a box around the
  $\rm{}^{13}CO$ cloud.} Note that these data are the only ones for
which we have used gas masses estimated from molecular line
emission. They are estimated from $^{13}\rm{CO}$ data assuming an
excitation temperature from the $^{12}\rm{CO}$ line emission and a
CO--to--$\rm{}H_2$ abundance ratio of $8\times{}10^{-5}$. The studies
further assume that the $^{12}\rm{CO}$--to--$^{13}\rm{CO}$ abundance
varies with galactocentric radius as described by
\citet{frerking1982:abundance}, respectively
\citet{milam2005:co-isotope-abundance}, depending on the study. We
stress that the $\rm{}^{13}CO$--based data are shown for completeness,
but that they suffer from other biases than the dust--based
observations that are the focus of this study. To calculate the
thermal velocity dispersion, a common gas temperature of $10~\rm{}K$
is assumed for all clouds in these samples.\label{app:data}

\paragraph{HMSF Clumps} We have compiled published data from
\citet{wienen2012:ammonia}, who present a comprehensive catalog of
cold (hence likely prestellar) high mass clumps identified from an
unbiased continuum survey of the inner Galactic Plane at
$870~\rm{}\mu{}m$. Masses are determined from the latter data.  We use
their $\rm{}NH_3$ (1,1) Effelsberg 100m--telescope data to obtain
velocity dispersions and estimate dust temperatures for mass
measurements. This allows to determine $\alpha$ for 260 clumps with
well--known distances: these clouds were either at tangential points,
or the distance ambiguity was previously resolved by
\citet{roman-duval2010:grs-clouds} on the basis of $21~\rm{}cm$
observations of H\textsc{i}.

\paragraph{non--HMSF Cores} Results on Perseus and the Pipe Nebula
represent non--HMSF clouds. We have included masses from
near--infrared extinction data \citep{lada2008:pipe-nature} and
kinematics from GBT $\rm{}NH_3$ (1,1) data
\citep{rathborne2008:pipe-nh3} for 29 cores in the Pipe Nebula. The
temperature is fixed to the average $\rm{}NH_3$--based temperature of
$10~\rm{}K$. In Perseus, we have combined masses from Bolocam dust
continuum data \citep{enoch2006:perseus} with gas temperatures and
velocity dispersions gleaned from GBT $\rm{}NH_3$ (1,1) data
\citep{rosolowsky2007:perseus_nh3}. \citeauthor{enoch2006:perseus}
provide the integrated flux density for several apertures. We chose to
adopt an aperture size of $40\arcsec$ to derive mass estimates, since
it closely matches the GBT \AMM\ beam.

\paragraph{HMSF Cores} Since HMSF clouds are typically more distant
than non--HMSF regions, interferometer observations are required to
achieve a spatial resolution similar to the one of observations of low
mass cores. We thus compile a large sample of high resolution
observations of potential prestellar stages of high mass star
formation. For the quiescent cores in Orion, we have calculated
$\alpha$ combining gas kinematics and temperatures from the VLA
$\rm{}NH_3$ (1,1) data of \citet{li2012:orion} with mass estimates
from SCUBA $850~\rm\mu{}m$ observations by
\citet{nutter2007:orion}. We have estimated $\alpha$ for the high mass
cores identified in G29.96$-$0.02 and W48 HMSF regions studied by
\citet{pillai2011:initial-conditions}. For this, we used the 3\,mm
dust continuum cores with associated {$\mathrm{NH_2D}$}
1$_{11}$--1$_{01}$ emission, and used these tracers to determine
masses and kinematics, respectively. The temperature is fixed to the
average $\rm{}NH_3$--based temperature of $16~\rm{}K$. To characterize
the \citet{sridharan2005:high-mass-starless-cores} sample, we glean
gas kinematics and estimated dust temperatures from the $\rm{}NH_3$
data presented by \citeauthor{sridharan2005:high-mass-starless-cores},
and then estimate masses from dust emission observations first
reported by \citet{beuther2002:hmpo-densities}. Data from a recent
ALMA study of high mass cores by \citet{tan2013:hmsc} are also
included. We used emission from dust and $\rm{}N_2D^+$ to constrain
masses and gas motions. Since \citeauthor{tan2013:hmsc} use a
different gas--to--dust ratio than us (i.e., 147 vs.\ our value of
100), we recompute the mass (given in their Table~3) for the same
gas--to--dust ratio as we have adopted in this work. We follow
\citeauthor{tan2013:hmsc} in assuming dust and gas temperatures of
$10~\rm{}K$.

\begin{table*}
\caption{Data Summary\label{tab:data-summary}}
\begin{center}
\begin{tabular}{lllllllll}
\hline \hline
Sample & $\langle R \rangle $ & sample median$^a$ $\langle \Sigma_M \rangle$ & SF Mode$^b$ &
$\sigma_v$ from & total mass from \\
      & pc  &  \gcmsq \\
\hline
Enoch et al.\ & 0.02  & 0.04   & non--HMSF      & \AMM\ (1,1)          & 1100 \micron\ \\
Heyer et al.\ & 3.91  & 0.04  & undetermined    & \tCO\ (1--0)                 & \tCO\ (1--0) \\
Lada et al.\ & 0.12  & 0.01  & non-HMSF      & \AMM\ (1,1) & NIR extinction \\
Li et al.\ & 0.04  & 0.18   & HMSF     & \AMM\ (1,1)                 & 850\micron \\
Pillai et al.\ & 0.29  & 0.12  & HMSF     & \DAMM\ (1$_{11}$--1$_{01}$)   & 3500 \micron \\
Roman--Duval et al.\ & 8.33  & 0.03   & undetermined    & \tCO\ (1--0) &  \tCO\ (1--0)  \\
Sridharan et al.\ & 0.22  & 0.10 & HMSF     & \AMM\ (1,1)                 & 1200\micron \\
Tan et al.\ & 0.06  & 0.13 & HMSF     & \NTD\ (3--2)                  & 1338\micron \\
Wienen et al.\ & 0.68  & 0.15 & HMSF     & \AMM\ (1,1)                 & 870\micron \\
\hline
\end{tabular}
\end{center}
\smallskip
$^a$the median value of $\langle \Sigma_M \rangle$
determined for a given sample\smallskip

$^b$mode of star formation (i.e., high--mass stars are present or
not, or should form in future), as judged by the original authors;
for ``undetermined'' samples, the star formation modes of the clouds
have not been assessed, but are likely mixed
\end{table*}

\section{Observed Trends in Virial Parameters}
\label{sec:trends}

\subsection{Observed Trends\label{sec:observed-trends}}
As seen in Figure~\ref{fig:vp-observed}, all of the data presented here
follow a number of common trends. Most fundamentally, within a
given sample, all data follow a power--law
\begin{equation}
\alpha=\alpha_0\cdot{}(M/10^3\,M_{\sun})^{h_{\alpha}}
\label{eq:vp-trend-1}
\end{equation}
with a similar slope, $h_{\alpha}$, and a range of intercepts,
$\alpha_0$. This power--law behavior has already been noticed by
\citetalias{bertoldi1992:pr_conf_cores} and also reported by, e.g.,
\citet{lada2008:pipe-nature} and \citet{foster2008:nh3-analysis}. To
illustrate these trends, fits to the \citet{lada2008:pipe-nature} and
\citet{wienen2012:ammonia} samples are drawn into
Fig.~\ref{fig:vp-observed}.

The sequences formed by a given sample terminate at a maximum mass,
$M_{\rm{}max}$, corresponding to a minimum virial parameter,
$\alpha_{\rm{}min}$. This is remarkable, since larger masses---and
therefore lower virial parameters---would be easily detected, if
present. By contrast, sequence terminations at lower mass cannot be
determined due to limited mass sensitivities. To highlight this trend,
one may rewrite Eq.~(\ref{eq:vp-trend-1}) as
\begin{equation}
\alpha=\alpha_{\rm{}min}\cdot{}(M/M_{\rm{}max})^{h_{\alpha}}\,{},
\label{eq:vp-trend-2}
\end{equation}
where
$\alpha_0=\alpha_{\rm{}min}\cdot{}(M_{\rm{}max}/10^3\,M_{\sun})^{-h_{\alpha}}$.
Table~\ref{tab:vp-laws} summarizes the power--laws representing the
various samples\footnote{In practice, we fit linear laws
  $y=\log_{10}(\alpha_0)+h_{\alpha}\cdot{}x$ to data of the form
  $x=\log_{10}(M/10^3\,M_{\sun})$ and $y=\log_{10}(\alpha)$.}. To
gauge the accuracy of these fits, the table also lists the root mean
square (RMS) deviation between the logarithms of the actual
observations and the fit,
$\delta_{\log_{10}(\alpha)}=\langle{}[\log_{10}(\alpha_i)-\log_{10}(\alpha(M_i))]^2\rangle^{1/2}$.

\begin{table}
\caption{Virial Parameter Power Laws\label{tab:vp-laws}}
\begin{center}
\begin{tabular}{llllllllll}
\hline \hline
Sample & $\alpha_0$ & $\alpha_{\rm{}min}$ & $M_{\rm{}max}/M_{\sun}$ &
$h_{\alpha}$ & $\delta_{\log_{10}(\alpha)}$ \\ \hline
Enoch et al.\ & 0.00 & 0.33 & 6 & $-$0.99 & 0.19\\
Heyer et al.\ & 3.11 & 1.71 & $1.4\times{}10^5$ & $-$0.12 & 0.30\\
Lada et al.\ & 0.05 & 0.65 & 20 & $-$0.68 & 0.15\\
Li et al.\ & 0.02 & 0.64 & 16 & $-$0.79 & 0.21\\
Pillai et al.\ & 0.17 & 0.15 & $1.3\times{}10^3$ & $-$0.61 & 0.19\\
Roman--Duval & 2.00 & 0.16 & $1.0\times{}10^6$ & $-$0.37 & 0.33\\
et al.\ \\
Sridharan & 0.55 & 1.11 & 226 & $-$0.47 & 0.13\\
et al.\ \\
Tan et al.\ & 0.06 & 0.39 & 43 & $-$0.62 & 0.17\\
Wienen et al.\ & 0.68 & 0.20 & $1.9\times{}10^4$ & $-$0.43 & 0.27\\ \hline
\end{tabular}
\end{center}
\end{table}

While some trends are similar for all samples, others differ among
the various studies.
\begin{enumerate}
\item When considering objects of increasing mass, all samples
  terminate at a maximum mass and minimum virial parameter,
  $M_{\rm{}max}$ and $\alpha_{\rm{}min}$. Values
  $\alpha_{\rm{}min}\ll{}2$ are observed for a number of samples. Some
  observations for individual fragments are even below the
  $\alpha_{\rm{}min}$ derived from fits to the samples.
\item All samples have very similar $\alpha(M)$ slopes,
  $h_{\alpha}$, observed to be in the range
  $0<-h_{\alpha}<1$.
\item The samples do significantly differ in their intercepts,
  $\alpha_0$. Differences by more than an order of magnitude are observed.
\end{enumerate}
Note that the observed anti--correlation between mass and virial
  parameter is not a consequence of an unfortunate combination of
  errors in mass estimates and the inherent relation
  $\alpha{}\propto{}M^{-1}$. Uncertainties in mass estimates are of
  order of a factor 2 (Sec.~\ref{sec:uncertainty-mass-estimates}),
  while all samples span more than an order of magnitude in
  mass. Errors in mass estimates can therefore not significantly
  affect the observed correlations.

\subsection{Virial Parameter Slope\label{sec:vp-slope}}
The trends in virial parameter slope and intercept result from the
slopes of the well--known mass--size and linewidth--size relations for
molecular clouds. To show this, we express the virial parameter slope
as $\d{}\log(\alpha)/\d{}\log(M)$, the mass--size slope as
$\d{}\log(M)/\d{}\log(R)$, and the linewidth--size relation as
$\d{}\log(\sigma_v)/\d{}\log(R)$. Logarithmic differentiation of the
definition of the virial parameter
(Eq.~\ref{eq:virial-parameter-definition}) yields
\begin{equation}
\frac{\d \log(\alpha)}{\d \log(M)} =
       \frac{
         2 \frac{\d \log(\sigma_v)}{\d \log(R)} + 1 -
         \frac{\d \log(M)}{\d \log(R)}
       }{
         \frac{\d{}\log(M)}{\d{}\log(R)}
       }
\label{eq:vp-slope-derivation}
\end{equation}
(see Appendix~\ref{app:vp-slope}). This demonstrates that the various
slopes directly depend on another. In practice, as seen in
Table~\ref{tab:vp-laws}, the virial parameter slope varies little
between the various samples. Following
Eq.~(\ref{eq:vp-slope-derivation}), this is a consequence of how
mass--size and linewidth--size laws combine in individual samples.

\begin{figure}
\includegraphics[width=\linewidth]{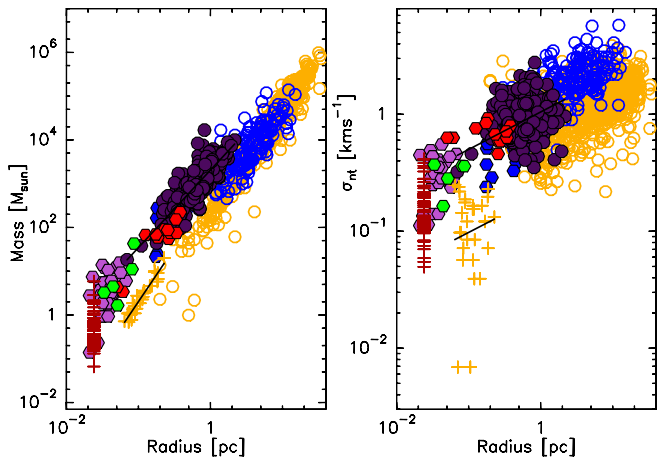}
\caption{Mass--size and linewidth--size trends in our
  catalogue. Colored symbols refer to the samples indicated in
  Fig.~\ref{fig:vp-observed}.\label{fig:larson-laws}}
\end{figure}

To explore this a bit more, Fig.~\ref{fig:larson-laws} shows the
connections between velocity dispersion, radius, and mass. This
representation suggests that the cloud fragments in a given sample
obey common mass--size and linewidth--size relations described by
power--laws (note that we investigate trends for the non--thermal
velocity dispersion, $\sigma_{v,{\rm{}nt}}$),
\begin{gather}
M = M_0\cdot{}(R/{\rm{}pc})^{h_M}\quad{\rm{}and}\label{eq:mass-size}\\
\sigma_{v,\rm{}nt} =
\sigma_{v,0}\cdot{}(R/{\rm{}pc})^{h_{\sigma_v}}\,{}.\label{eq:linewidh-size}
\end{gather}
Table~\ref{tab:mass-linewidth-size} lists fitted\footnote{Again, we
  employ linear fits to properties $y=\log_{10}(M)$, respectively
  $y=\log_{10}(\sigma_v)$, and $x=\log_{10}(R)$.} properties for the
various samples. To indicate trends, the fits for the
\citet{lada2008:pipe-nature} and \citet{wienen2012:ammonia} samples
are indicated in Fig.~\ref{fig:larson-laws}.

For linewidth--size laws, we experiment with
assuming a common slope of $h_{\sigma_v}=0.32$. This is the slope
derived when fitting all data with a common relation. The fit gives an
intercept $\sigma_{v,0}=0.8~\rm{}km\,s^{-1}$. The RMS deviation
between fit and observations is
$\delta_{\log_{10}(\sigma_v)}=\langle{}[\log_{10}(\sigma_{v,i}/\sigma_v(R_i))]^2\rangle^{1/2}=0.24$,
i.e., the RMS scatter is of the order of a factor
$10^{0.24}=1.7$. When we use $h_{\sigma_v}=0.32$ to fit individual
samples, the fit results and deviations given in
Table~\ref{tab:mass-linewidth-size} are obtained. For all but one
sample, we find good fits characterized by
$\delta_{\log_{10}(\sigma_v)}<0.2$, i.e., an RMS scatter of a factor
$<1.6$. For the \citet{lada2008:pipe-nature} study,
$\delta_{\log_{10}(\sigma_v)}=0.35$, equivalent to a factor
$10^{0.35}=2.2$, is obtained. As already noted by
\citeauthor{lada2008:pipe-nature}, the latter probably reflects the
fact that the dense cores in the Pipe Nebula are ``coherent''
(\citealt{barranco1998:coherent_cores},
\citealt{goodman1998:coherent_cores}), i.e., thermal motions play a
significant or dominant role. The residual non--thermal motions are
often negligible, and they do not closely obey linewidth--size
laws. With the exception of the \citeauthor{lada2008:pipe-nature}
sample, the derived intercepts are within a factor 2 from the
intercept derived for a common fit to all data,
$\sigma_{v,0}=0.8~\rm{}km\,s^{-1}$.

This shows that the mass--size and linewidth--size laws are indeed
relatively similar for all samples. One could now use the
approximations $\d{}\log(M)/\d{}\log(R)\approx{}h_M$ and
$\d{}\log(\sigma_v)/\d{}\log(R)\approx{}h_{\sigma_v}$ to derive
$h_{\alpha}$ as $\d{}\log(\alpha)/\d{}\log(M)$ via
Eq.~(\ref{eq:vp-slope-derivation}). In practice, though, this is an
exercise of limited value, given the approximations involved. In any
case, as shown by Eq.~(\ref{eq:vp-slope-derivation}), the slopes
depend on another.\medskip

We note that a virial parameter slope $h_{\alpha}=-2/3$ has previously
been predicted by \citetalias{bertoldi1992:pr_conf_cores}. For
example, \citet{lada2008:pipe-nature} interpret their data in
\citetalias{bertoldi1992:pr_conf_cores}'s framework. Specifically, a
value $h_{\alpha}=-2/3$ is expected if all fragments in a sample have
the same mean density and the same velocity dispersion. To see this,
one may use the mean density
$\langle{}\varrho{}\rangle=M/(4/3\,\pi\,R^3)$ to rewrite the virial
parameter as
$\alpha=C\cdot{}\sigma_v^2\,M^{-2/3}\,\langle{}\varrho{}\rangle^{-1/3}$,
where $C$ is a numerical constant. This implies
$\alpha\propto{}M^{-2/3}$, and thus $h_{\alpha}=-2/3$, if $\sigma_v$
and $\langle{}\varrho{}\rangle$ are
constant. \citetalias{bertoldi1992:pr_conf_cores} highlight that
$\langle{}\varrho{}\rangle$ is indeed constant in samples of
pressure--confined fragments---i.e., where self--gravity is
negligible---that are subject to the same confining pressure and have
a common $\sigma_v$. Many of the slopes noted in
Table~\ref{tab:vp-laws} are about consistent with $h_{\alpha}=-2/3$
and fulfil this prediction for pressure--confined fragments. Note,
however, that self--gravity plays a significant role in most of the
samples, as indicated by values of $\alpha_{\rm{}min}$ much below
$\alpha_{\rm{}BE}$. The aforementioned theory predicting
$h_{\alpha}=-2/3$, which only applies to fragments with negligible
self--gravity, does thus not apply.

\begin{table}
\caption{Mass--Size and Linewidth--Size Laws\label{tab:mass-linewidth-size}}
\begin{center}
\begin{tabular}{lllllllllllll}
\hline \hline
Sample & $h_M$ &
$\sigma_{v,0}/\rm{}km\,s^{-1}$ &
$\delta_{\log_{10}(\sigma_v)}$\\
\hline
Enoch et al.\ & ---$^a$ & 0.49 & 0.20\\
Heyer et al.\ & $1.40\pm{}0.07$ & 1.42 & 0.16\\
Lada et al.\ & $2.46\pm{}0.14$ & 0.20 & 0.35\\
Li et al.\ & $1.86\pm{}0.66$ & 0.85 & 0.18\\
Pillai et al.\ & $2.22\pm{}0.61$ & 0.69 & 0.15\\
Roman--Duval et al.\ & $2.36\pm{}0.02$ & 0.69 & 0.18\\
Sridharan et al.\ & $1.80\pm{}0.23$ & 1.15 & 0.12\\
Tan et al.\ & $2.96\pm{}0.95$ & 0.74 & 0.12\\
Wienen et al.\ & $1.77\pm{}0.07$ & 1.05 & 0.16\\
\hline
\end{tabular}
\end{center}
\smallskip
$^a$no trend with radius, since all observations are obtained for a
fixed aperture
\end{table}

\subsection{Virial Parameter Intercept}
To explore the virial parameter intercepts, consider mean mass surface
densities, $\langle{}\Sigma_M\rangle{}=M/(\pi{}R^2)$. Substitution of
the linewidth--size law in Eq.~(\ref{eq:virial-parameter}), plus
replacing $R$ with $\langle{}\Sigma_M\rangle$ and $M$, yields
\begin{equation}
\begin{array}{l}
\displaystyle
\alpha=1.2\,\left(\frac{\sigma_v}{\sigma_{v,\rm{}nt}}\right)^2
  \\
\displaystyle
  \quad\quad\,\, \cdot \left(\frac{\sigma_{v,0}}{\rm{}km\,s^{-1}}\right)^2
\frac{
  (M/10^3\,M_{\sun})^{h_{\sigma_v}-1/2}
}{
  (\langle{}\Sigma_M\rangle{}/0.066~{\rm{}g\,cm^{-2}})^{h_{\sigma_v}+1/2}
}\,,
\end{array}
\label{eq:virial-parameter-intercepts}
\end{equation}
where we use that
$1~{\rm{}g\,cm^{-2}}=4800\,M_{\sun}\,\rm{}pc^{-2}$. The factor
$(\sigma_v/\sigma_{v,\rm{}nt})^2$ provides a correction in the case
that non--thermal gas motions are important. Note that
Eq.~(\ref{eq:virial-parameter-intercepts}) does not present an
approximation; for example, provided parameters for individual
fragments are substituted, Fig.~\ref{fig:vp-observed} could be
constructed via Eq.~(\ref{eq:virial-parameter-intercepts}) in all
details.

This shows that the virial parameter intercept strongly depends on the
linewidth--size intercept and the mean mass surface density of a
sample. The latter varies dramatically between the samples, as seen in
Table~\ref{tab:data-summary}, largely due to the sensitivity of
different observational methods. As in the case of
Eq.~(\ref{eq:vp-slope-derivation}), we abstain from attempt to
substitute characteristic values for samples into
Eq.~(\ref{eq:virial-parameter-intercepts}). Experimentation shows
that, e.g., $\sigma_{v,0}$ and $\langle{}\Sigma_M\rangle$ vary too
strongly within samples to derive meaningful results.

\subsection{Low Virial Parameters}
To fully describe the observed virial parameter laws, we finally need
to interpret the observed minimum virial parameters
$\alpha_{\rm{}min}\ll{}2$. The remainder of the paper is dedicated to this.

\section{Implications of Low Virial Parameters}
Figure~\ref{fig:vp-observed} and Table~\ref{tab:vp-laws} show that the
observed minimum virial parameters in many samples fall below the
fiducial critical value $\alpha{}=2$ by a factor of 5 or more. This is
a significant difference beyond the range expected from observational
uncertainties, as we explore in
Sec.~\ref{sec:vp_correct-relevant}. Given critical virial parameters
$\alpha_{B=0}\gtrsim{}2$, fragments characterized by $\alpha{}\ll{}2$
must be unstable to collapse unless they are supported by significant
magnetic fields. In this section, we discuss what this conclusion
means for our understanding of the star formation
process.\label{sec:sf-physics}

To make the discussion more readable, some of the quantitative details
of the discussion have been removed to
Appendix~\ref{app:implications}. The current section focuses on the
main implications from the analysis.

\begin{figure*}
\includegraphics[width=\linewidth]{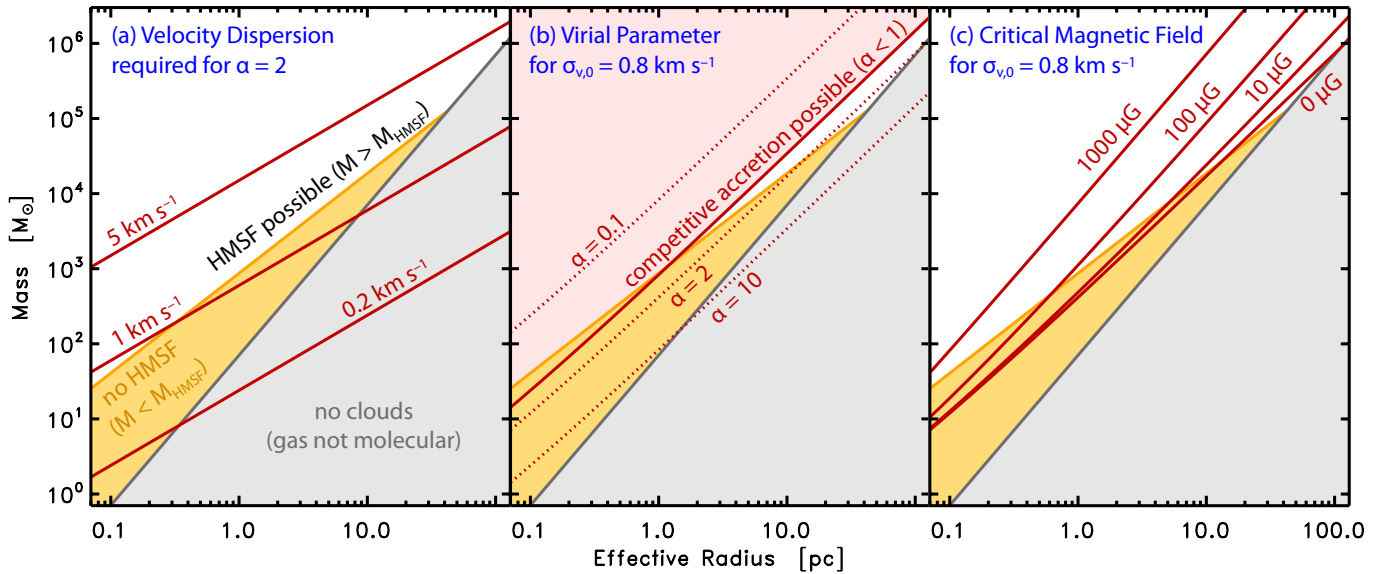}
\caption{Implications for star--formation physics. Yellow shading
  highlights the region in which cloud fragments are not massive
  enough to host high mass star formation (HMSF). The limit drawn in
  this figure is taken from \citet{jenthu2010:irdcs} and assumes---in
  contrast to all other parts of the paper---a dust opacity
  $\kappa(\nu)=\kappa_{\rm{}OH}(\nu)/1.5$. Red lines in panel~(a) give
  the minimum velocity dispersion needed to render a non--magnetized
  cloud fragment stable against gravitational collapse
  ($\alpha=2$). Panel~(b) presents estimated virial parameters as a
  function of mass and size, as expected for $15~\rm{}K$ gas
  temperature and a common linewidth--size law with
  $h_{\sigma_v}=0.32$ and $\sigma_{v,0}=0.8~\rm{}km\,s^{-1}$. Red
  shading indicates where competitive accretion is expected to be
  possible ($\alpha<1$). In panel~(c) the same temperature and
  linewidth--size law is assumed to estimate the minimum ``critical''
  magnetic field strength needed to render a cloud fragment stable
  against gravitational collapse. The grey shading indicates cloud
  fragments with a mean visual extinction $A_V<1~\rm{}mag$. That part
  of the parameter space is devoid of molecular clouds, since the mean
  visual extinction is too low to shield the molecular gas from
  photodissociation.\label{fig:physics}}
\end{figure*}

Note that the smallest virial parameters are found in regions of high
mass star formation (HMSF). See for example
Fig.~\ref{fig:vp-observed}, where HMSF regions are indicated by filled
symbols. Specifically, the HMSF samples from
\citet{pillai2011:initial-conditions}, \citet{li2012:orion},
\citet{tan2013:hmsc}, and \citet{wienen2012:ammonia} contain virial
parameters much smaller than those in the non--HMSF samples by
\citet{lada2008:pipe-nature} and \citet{enoch2006:perseus}. Also
Table~\ref{tab:vp-laws} lists the smallest $\alpha_{\rm{}min}$ for
HMSF sites. This is a consequence of the high mass of HMSF clouds,
which are observed to exceed an approximate size--dependent threshold,
$M(R)>M_{\rm{}HMSF}(R)$ where
\begin{equation}
M_{\rm{}HMSF}(R) = 870\,M_{\sun} \,
  \left( \frac{\kappa(\nu)}{\kappa_{\rm{}OH}(\nu)/1.5} \right)^{-1}
  \left( \frac{R}{\rm{}pc} \right)^{1.33}
  \label{eq:threshold-hmsf}
\end{equation}
(\citealt{jenthu2010:irdcs}; see yellow shading in
Fig.~\ref{fig:physics}). This relation depends on the adopted dust
opacity law. A relation $\kappa(\nu)=\kappa_{\rm{}OH}(\nu)/1.5$ was
adopted in the original work (\citealt{kauffmann2010:mass-size-i,
  kauffmann2010:mass-size-ii};
\citealt{jenthu2010:irdcs}). Equation~(\ref{eq:low-alpha-hmsf})
demonstrates that $M(R)>M_{\rm{}HMSF}(R)$ implies
$\alpha{}\lesssim{}1$ for radii $\sim{}0.1~\rm{}pc$.

In essence, the observed virial parameters are low because the
observed velocity dispersions are low for the given mass and size of a
fragment. To provide a reference, Fig.~\ref{fig:physics}(a)
illustrates the velocity dispersion needed to achieve $\alpha=2$.

\subsection{Short Lifetimes of High--Mass Starless
  Cores\label{sec:absence-starless-cores}}
The study of high mass star formation is still trying to answer a
central and important riddle: why are there no starless cores of high
mass and density? The absence of such cores is significant, given that
many starless cores exist in non--HMSF regions (see, e.g.,
\citealt{evans2009:c2d-summary} for statistics). After the discovery
of Infrared Dark Clouds (IRDCs), it has been speculated that some
cores within these clouds would represent high mass starless cores
(e.g., \citealt{egan1998:irdcs},
\citealt{carey1998:irdc-properties}). Follow--up studies do indeed
show that IRDCs form high--mass stars (e.g.,
\citealt{beuther2005:hmsf-onset}, \citealt{pillai2006:g11},
\citealt{rathborne2007:irdc-msf}). Such work also revealed potential
high mass starless cores in IRDCs (e.g.,
\citealt{sridharan2005:high-mass-starless-cores},
\citealt{swift2009:starless-irdc}
\citealt{pillai2011:initial-conditions} \citealt{wienen2012:ammonia},
\citealt{tan2013:hmsc}). However, to our best knowledge, the few
objects that were studied with targeted follow--up observations
are not found to be clear--cut high--mass starless
cores\footnote{\citet{zhang2011:tale-two-cores} find $\rm{}H_2O$
  masers near the candidate HMSF starless core from
  \citet{swift2009:starless-irdc}, and \citet{wang2006:irdc-masers}
  find them in the region studied by \citet{tan2013:hmsc}. These
  masers originate in the outflows from young stars, and their
  presence indicates the existence of such stars in these clouds. It
  is, however, not clear in which of the many cores in the region
  these stars do form. Further, at higher spatial resolution,
  \citet{zhang2011:tale-two-cores} find no compact cores of high mass
  in the \citet{swift2009:starless-irdc} candidate.}.

This implies that high--mass starless cores are rare. Such
a trend further implies that the lifetime of such cores must be
low. The low observed virial parameters might explain why this is the
case.

Cloud fragments with virial parameters $\alpha{}\ll{}2$ might collapse
very quickly, essentially in a free--fall time, if magnetic fields are
insignificant for support against self--gravity. This follows from the
fact that non--magnetized cloud fragments with virial parameters
$\alpha{}\ll{}\alpha_{\rm{}BE}$ are not supported against
gravitational collapse. In HMSF regions, mass and size are related by
the approximate threshold for high--mass star formation,
Eq.~(\ref{eq:threshold-hmsf}). When we substitute this relation into
the equation for the free--fall timescale
(Appendix~\ref{app-sec:absence-starless-cores}), we find that
$\tau_{\rm{}ff}<5.5\times{}10^4~{\rm{}yr}\cdot{}(M/10\,M_{\sun})^{0.62}$
holds for cloud fragments $M>M_{\rm{}HMSF}$ deemed able to form
high--mass stars. This implies short lifetimes for non--magnetized
fragments, just as needed to explain the absence of high--mass
starless cores.

\subsection{``Turbulent Core Accretion'' in
  HMSF\label{sec:turbulent-core-accretion}}
High--mass star formation requires high accretion rates onto the
stars. For example, in \citet{pillai2011:initial-conditions} we
summarize previous work suggesting that accretion rates
$\ge{}10^{-4}\,M_{\sun}\,\rm{}yr^{-1}$ are needed to form stars of a
mass $\ge{}10\,M_{\sun}$ during the estimated duration of the
accretion phase $\le{}10^5~\rm{}yr$. Accretion rates of this order
are, for example, expected for the collapse of spheres that were
initially in hydrostatic equilibrium and supported by
\emph{sufficiently fast} random gas motions. In non--magnetized
spheres, stellar accretion rates
$\dot{M}_{\star}\le{}2.3\times{}10^{-4}\,M_{\sun}\,{\rm{}yr}\cdot{}(\sigma_v/{\rm{}km\,s^{-1}})^3$
are then possible \citep{shu1977:self-sim_collapse,
  mckee2002:massive-sf, mckee2003:massive-sf,
  pillai2011:initial-conditions}. The non--magnetized version of
  the ``turbulent core'' HMSF model by \citet{mckee2002:massive-sf,
  mckee2003:massive-sf} posits that high accretion rates in HMSF do
indeed result from the collapse of cloud fragments initially supported
by high velocity dispersions. However, there are issues with this
picture.

For example, as we show throughout this paper, many HMSF cloud
fragments have virial parameters much below the critical value for
non--magnetized media. This means that many HMSF fragments are not in
hydrostatic equilibrium, and so the original model from
\citet{mckee2002:massive-sf} does not apply, as it requires
$\alpha{}\gtrsim{}2$. This problem is also noted by
\citet{tan2013:hmsc}. They suggest to include additional support from
magnetic fields. This would, however, reduce the role of $\sigma_v$
and represent a major modification of the initial
\citet{mckee2002:massive-sf} model. Such refinements are explored by
\citet{mckee2003:massive-sf}. In this more complete picture, the
velocity dispersion $\sigma_v$ is not the only factor controlling
$\dot{M}_{\star}$.

Another issue is that, for the non--magnetized case, the substitution
of observed velocity dispersions yields accretion rates only
marginally consistent with
$\dot{M}_{\star}\ge{}10^{-4}\,M_{\sun}\,\rm{}yr^{-1}$
\citep{pillai2011:initial-conditions}. If magnetic fields are present,
the model accretion rates would be higher by a factor 6.6 when
adopting the fiducial magnetic field properties proposed by
\citeauthor{mckee2003:massive-sf} (\citeyear{mckee2003:massive-sf};
their $H_0=1$ and $\phi_B=2.8$).

\subsection{Competitive Accretion\label{sec:competitive-accretion}}
It is currently not clear by which accretion mechanism high--mass
stars grow in mass. One theory posits that individual dense cores
produce individual stars or close binaries. Here, we refer to this
process as ``monolithic accretion''
(\citealt{zinnecker2007:msf-review}; e.g., the
  \citealt{mckee2002:massive-sf, mckee2003:massive-sf} models belong
  to this category).  Alternative theories propose that no
well--defined HMSF dense cores exist. In this scenario of
``competitive accretion'', several stars vie with another to accrete
from a common gas reservoir. This star formation mechanism has been
studied by, e.g., \citet{bonnell1997:mass-spectrum,
  bonnell2001:competitive-accretion, bonnell2001:accretion-imf} and
\citet{bate2005:imf}.

\citet{krumholz2005:non-competitive} show that competitive accretion
requires virial parameters $<1$. This follows from the constraints
that competitive accretion works best if the gas densities are high
and the relative velocity between star and gas are small, and can be
evaluated for the case of Bondy--Hoyle accretion.  In their original
work, \citet{krumholz2005:non-competitive} compiled virial parameters
from a small sample of clouds and clumps. Those data suggested that
$\alpha{}\gtrsim{}1$. \citeauthor{krumholz2005:non-competitive}
interpreted this as evidence that competitive accretion is not
possible in HMSF regions.

The larger sample depicted in Fig.~\ref{fig:vp-observed}, however,
shows that $\alpha{}<1$ is \emph{frequently} observed in regions of
high mass star formation. This new result demonstrates that
competitive accretion would be possible in many of the clouds studied
here. Note that this does \emph{not}, however, constitute evidence
against monolithic accretion. We also caution that some of the
fragments with $\alpha<1$ might not be massive enough to form entire
clusters. In that case, competitive accretion cannot operate, since it
requires the presence of a significant
cluster. Figure~\ref{fig:physics}(b) illustrates the mass--size domain
in which $\alpha<1$, so that competitive accretion becomes possible.

Note, further, that \citet{krumholz2005:non-competitive} propose the
additional limit that competitive accretion requires
$\alpha^2\,M\lesssim{}10\,M_{\sun}$. The exact value of this threshold
depends on model details \citep{krumholz2005:non-competitive}, and it
may be larger than the fiducial value by an order of magnitude
\citep{bonnell2006:competitive-accretion}. When selecting fragments
with $M(R)>M_{\rm{}HMSF}(R)$ from the HMSF samples, 6 objects fulfil
$\alpha^2\,M<10\,M_{\sun}$.

\subsection{Fragments with Small Virial Parameters\\ are not
  Collapsing\label{sec:no-collapse}}
By definition, $\alpha<\alpha_{\rm{}cr}$ implies that a cloud fragment
is susceptible to collapse. Thus, it has often been argued that cloud
fragments with virial parameters
$\alpha\ll{}\alpha_{\rm{}BE}\approx{}2$ are indeed collapsing to form
a star. However, this is a flawed argument: provided energy is
conserved, fragments well into collapse contain gas rapidly moving
inward, and velocity dispersions obtained under these conditions imply
virial parameters $\approx{}2a=4\pm{}2$ (see
  Appendix~\ref{app-sec:no-collapse}), since $a=2\pm{}1$. As a
consequence, fragments with $\alpha\ll{}\alpha_{\rm{}BE}\approx{}2$
are \emph{not} likely to be in a state of collapse. This has
previously been realized by, for example,
\citet{larson1981:linewidth_size} and
\citet{ballesteros-paredes2006:six-myths}.

One important caveat is that this argument assumes conservation of
energy. This is a reasonable ansatz. To circumvent this constraint,
energy would need to be drained from the system at a very high
rate. Doing so is not a trivial problem (see
Appendix~\ref{app-sec:no-collapse} for some example calculations).

It thus appears plausible that fragments with virial parameters
$\ll{}2$ are \emph{not} collapsing. Fragments in this state would then
need to be supported by additional forces. Alternatively, they might
be short--lived and soon collapse in a free--fall time
(Sec.~\ref{sec:absence-starless-cores}).

\subsection{Evidence for Significant Magnetic
  Fields?\label{sec:magnetic-fields}}
Let us assume that cloud fragments with $\alpha\ll{}2$ are indeed not
undergoing collapse, as argued in
  Sec.~\ref{sec:no-collapse}. The most straightforward explanation
for such fragments would be that they are supported against collapse
by significant magnetic fields.

Equation~(\ref{eq:alpha_cr-magnetic-approx}) gives the critical virial
parameter in this situation. This relation only depends on
$M_{\Phi}/M_{\rm{}BE}$, which essentially measures the relative
importance of support from magnetic fields relative to random gas
motions. If we require that $\alpha\ge{}\alpha_{\rm{}cr,\Phi}$, we can
rewrite Eq.~(\ref{eq:alpha_cr-magnetic-approx}) to obtain
\begin{equation}
\frac{M_{\Phi}}{M_{\rm{}BE}} \gtrsim \frac{2}{\alpha}-1 \, .
\label{eq:magn-virial-parameter}
\end{equation}
Combination of Eqs.~(\ref{eq:be-mass}, \ref{eq:flux-mass}) permits
us to estimate the magnetic field strength as
\begin{equation}
B = 81~{\rm{}\mu{}G} \, \frac{M_{\Phi}}{M_{\rm{}BE}} \,
  \left( \frac{\sigma_v}{\rm{}km\,s^{-1}} \right)^2
  \left( \frac{R}{\rm{}pc} \right)^{-1} \, .
\label{eq:magnetic-field-estimation}
\end{equation}
For example, the lowest observed virial parameters are of order
0.2. This implies mass ratios $M_{\Phi}/M_{\rm{}BE}\approx{}10$. Those
cloud fragments are also observed to have velocity dispersions
$\sim{}0.5~\rm{}km\,s^{-1}$, radii $\sim{}0.1~\rm{}pc$, and masses
$\sim{}10^2\,M_{\sun}$. Following
Eq.~(\ref{eq:magnetic-field-estimation}), this implies field strengths
$\sim{}2~\rm{}mG$. This is a very high value. However, it is
marginally in the range expected for cloud fragments of very high
density. \citet{crutcher2012:review} suggests an approximate
density--dependent upper limit to the magnetic field strength
(Appendix~\ref{app-sec:magnetic-fields}). Substitution of the
aforementioned masses and sizes gives field strengths
$B\lesssim{}1.5~\rm{}mG$, marginally consistent with the above
estimate $\sim{}2~\rm{}mG$.

Representative values for the maximum critical mass of magnetized
clouds, $M_B\lesssim{}M_{\rm{}BE}+M_{\Phi}$, are given in
Fig.~\ref{fig:physics}(c). All cloud fragments below the $B=0$ curve
are subcritical even if not supported by magnetic fields.

\subsection{Mass Estimates from Virial
  Analysis\label{sec:virial-mass}}
In the absence of better tracers, the virial mass
$M_{\rm{}vir}\equiv5\sigma_v^2R/G$ is often used to estimate the true
mass, $M$. Basically, this assumes that
$M\approx{}M_{\rm{}vir}$. Since $\alpha{}\equiv{}M_{\rm{}vir}/M$ by
definition, this requires that
$\alpha{}\approx{}1$. Figure~\ref{fig:vp-observed} shows that, for
individual cloud fragments, this assumption is violated by factors 10
and more: in this case, the mass estimated via a virial analysis is
off by the same factor. For individual objects, this large uncertainty
must be kept in mind when using masses estimated in this fashion.

That said, it might be appropriate to determine, e.g., the total gas
mass contained in a cloud sample via a virial analysis. As seen in
Fig.~\ref{fig:vp-observed}, the scatter is large for individual
objects, but the average virial parameter is often well defined for a
given sample, and is often observed to be of order unity. A careful
analysis of cloud samples could exploit such trends.

\section{Virial Parameters: Impact of Uncertainties}
The analysis above builds on various assumptions to calculate $\alpha$
and $\alpha_{\rm{}cr}$. Here we explore whether uncertainties in these
assumptions might affect our
conclusions.\label{sec:vp_correct-relevant}

\subsection{Observational Approaches}
The way in which properties are calculated can significantly affect
the results. This is seen in Fig.~\ref{fig:vp-observed}, where the
results do significantly differ between
\citet{heyer2009:scaling-relations} and
\citet{roman-duval2010:grs-clouds}, although both studies build on the
same set of observations. The difference lies in the way cloud
structure was extracted: \citeauthor{heyer2009:scaling-relations}
examine boxes drawn around clouds, which serves their purpose but is
not ideal to measure $\alpha$, while
\citeauthor{roman-duval2010:grs-clouds} closely follow the cloud shape
and derive properties better suited for a virial analysis. This
example demonstrates that determining the virial parameter is a
difficult task.

\subsection{Errors in Mass
  Estimates\label{sec:uncertainty-mass-estimates}}
The most commonly adopted dust model for cold dense cores, which is
also used in this paper, has been forwarded by
\citet{ossenkopf1994:opacities}. The resulting opacity is intermediate
to the values proposed by, e.g., \citet{kruegel1994:opacities} and
\citet{motte1998:ophiuchus}, who forward values higher or lower by a
factor 2. If we confine ourselves to the range in these models, even
if the true dust opacities were higher, that would account for only a
systematic uncertainty of a factor 2 in dust masses and therefore
virial parameters. An error in the opacity by a
factor $>10$, as needed to render fragments with estimated
$\alpha{}=0.2$ subcritical, seems implausible.

The mass is approximately inversely proportional to the adopted dust
temperature. For some objects in our analysis, dust temperatures were
estimated by assuming that they are similar to gas temperatures. For
example, using observations of $\rm{}NH_3$,
\citet{rathborne2008:pipe-nh3} derive temperatures for the sample of
\citet{lada2008:pipe-nature}, and \citet{li2012:orion} obtain own data
for their sample. For starless cores, such work typically finds
temperatures in the range 10--15$~\rm{}K$. Direct estimates of dust
temperatures from continuum emission yield similar values for regions
not in the immediate vicinity of HMSF (e.g.,
\citealt{hill2011:vela-c}). Following
\citet{kauffmann2008:mambo-spitzer}, the conversion factor between
dust emission and mass varies by a factor $\le{}1.75$ for dust
temperatures in the range 10--15$~\rm{}K$ and dust emission
observations at wavelengths $\ge{}850~\rm{}\mu{}m$ relevant for our
analysis. At higher temperatures, for example close to young emedded
stars, a temperature variation by $5~\rm{}K$ would result in smaller
uncertainties in the conversion factor. A factor of 1.75 thus provides
a conservative upper limit to the temperature--related uncertainties
in the virial parameter estimates presented here. In our analysis, we
adopt the dust temperature the original authors assumed based on their
in--depth knowledge of the observed region, unless better constraints
from complementary observations exist. We thus expect a much smaller
temperature and mass uncertainty to hold for most of the samples.

\subsection{Errors in Distance
  Estimates\label{sec:uncertainty-distance-estimates}}
Measurements of mass and radius scale with distance as $M\propto{}d^2$
and $R\propto{}d$, implying $\alpha{}\propto{}d^{-1}$. Distances for
the majority of star--forming regions are still derived from the
kinematics, based on a Galactic rotation curve. Accurate parallax
distance measurements of star forming regions recently revealed that
their motions deviate from circular motions by up to
$\approx{}15~\rm{}km\,s^{-1}$. The kinematic distances are generally
larger than parallax--based ones, and can be larger by up to a factor
two \citep{reid2009:parallaxes}. If this situation holds generally,
the virial parameter would be underestimated by up to the same factor
$\lesssim{}2$.

\subsection{Extreme Fragment
  Geometries \& Cloud Complexes\label{sec:uncertainty-fragment-geometries}}
The issue of non--spherical fragment geometries was already considered
in Sec.~\ref{sec:vp-summary}, based on
Appendix~\ref{app:vp-calculations}. We concluded that the virial
parameter is a good measure of the kinetic--to--potential energy
ratio, even when considering highly flattened or elongated structures
for a range of density gradients. It seems implausible that extreme
viewing geometries or density gradients could produce observed viral
parameters $\ll{}2$ in objects that are stable against collapse. Also
note that \emph{all} of the HMSF samples presented in
Fig.~\ref{fig:vp-observed} have minimum virial parameters
$\ll{}2$. The prevalence of small virial parameters strongly argues
against an influence of unusual viewing geometries.

Notice, however, that all the models discussed in
Sec.~\ref{sec:vp-summary} and Appendix~\ref{app:vp-calculations}
assume flat or centrally condensed density structures that roughly
obey a point symmetry. This assumption may be significantly violated
when considering larger complexes. When considered at their full
extent, regions like the Taurus molecular cloud (e.g.,
\citealt{goldsmith2008:taurus-fcrao}) are clearly not centrally
condensed and do not exhibit any clear symmetry. In fact, regions of
high gas density are distributed within large areas devoid of any
significant mass reservoir. It is not clear how well the gravitational
potential energy is approximated by the formalism adopted above. Also,
it is not clear that pressure from kinetic gas motions, magnetic
fields, etc., is really relevant for cloud support.

\subsection{Role of Rotation \& Surface
  Pressure\label{sec:uncertainty-rotation-pressure}}
\citet{caselli2002:n2h+} study rotation in a sample of dense cores of
relatively low mass that are located in the solar neighborhood (most
cores are closer than $500~\rm{}pc$; also see
\citealt{goodman1993:velocity-gradients}). They derive
rotational--to--gravitational energy ratios $<0.1$. This implies that
rotation can be neglected in these cores. Based on $8~\rm{}\mu{}m$
extinction mapping and Ammonia data from the VLA,
\citet{ragan2009:irdc-extinction, ragan2012:irdcs-vla} obtain virial
parameters of 0.1--0.7 for a sample of 6 IRDCs (their definition of
$\alpha$ is identical to ours). Even in this sample with $\alpha<1$,
\citet{ragan2012:irdcs-vla} demonstrate that the
rotational--to--gravitational energy ratios is
$\le{}2\times{}10^{-4}$. Again, rotation appears to be
negligible. However, we caution that the study of velocity gradients
in clouds is not a settled topic yet (e.g.,
\citealt{kirk2010:dense-core-dynamics-ii}).\medskip

The cloud models employed to gauge $\alpha_{\rm{}cr}$ assume that
cloud fragments are in equilibrium with a constant surface
pressure. Given the dynamic nature of clouds, this assumption can only
be fulfilled in an approximate sense, and might be significantly
violated in some situations. This can influence the stability of cloud
fragments.

However, deviations from pressure equilibrium, as well as fluctuations
in the external pressure, should only have a significant impact on
cloud fragments for which confinement by an outside pressure is
relevant. As discussed in Sec.~\ref{sec:vp-summary}, such cloud
fragments have virial parameters significantly larger than
$2a=4\pm{}2$. In other words, the external pressure plays no
significant role for the fragments with $\alpha{}\ll{}2$ that are the
focus of the current paper: these fragments are largely confined by
their own gravity.

\section{Summary}
The stability of cloud fragments against collapse is of utmost
importance for the study of molecular clouds and the star formation
process. It can be gauged using the virial parameter,
$\alpha\equiv{}5\sigma_v^2R/(GM)$, which is easily calculated from a
fragment's velocity dispersion, radius, and mass. Fragments are
unstable to collapse if they have supercritical virial parameters,
$\alpha<\alpha_{\rm{}cr}$. Non--magnetized cloud fragments are
expected to have critical virial parameters
$\alpha_{\rm{}cr}\gtrsim{}2$, while $\alpha_{\rm{}cr}\ll{}2$ is
possible for strong magnetic fields
(Sec.~\ref{sec:vp-summary}).\label{sec:summary}

Past research suggested that virial parameters $\alpha{}\gtrsim{}2$
prevail in clouds. This would suggest that collapse towards star
formation is a gradual and relatively slow process. It also suggests
that magnetic fields are not needed to explain the observed cloud
structure. But many recent studies, in particular of regions of
high--mass star formation (HMSF), yield much lower virial parameters
$\alpha<2$. To study the implications of these results, we compile a
catalogue of 1325 virial parameter estimates that are obtained in a
standardized fashion (Sec.~\ref{sec:analysis}). In particular, we
adopt a single dust opacity law for all samples and standardize the
definitions of $\alpha$, $M$, $R$, and $\sigma_v$. The results are
shown in Fig.~\ref{fig:vp-observed} and Table~\ref{tab:vp-laws}. We
find that, within a given sample, the virial parameters follow a trend
$\alpha=\alpha_{\rm{}min}\cdot{}(M/10^3\,M_{\rm{}max})^{h_{\alpha}}$
with $0<-h_{\alpha}<1$ (Sec.~\ref{sec:trends}). For many samples,
$\alpha_{\rm{}min}\ll{}2$.

These observations have a number of important implications for the
physics governing the formation of stars, in particular those of a
very high mass.
\begin{itemize}
\item The scarcity of high--mass starless cores could be
  explained (Sec.~\ref{sec:absence-starless-cores}). If such cores are
  not at all supported against collapse, as indicated by
  $\alpha{}\ll{}2$, they would collapse in a free--fall timescale
  $<10^5~\rm{}yr$. The starless phase would be shorter than this
  timescale and few high-mass cores should exist in this state.
\item ``Turbulent core accretion'' in HMSF (e.g.,
  \citealt{mckee2002:massive-sf, mckee2003:massive-sf}) would not be
  applicable, unless additional support comes from magnetic fields
  (Sec.~\ref{sec:turbulent-core-accretion}). For non--magnetized
    clouds, that model requires that random ``turbulent'' gas motions
  provide significant support against self--gravity. This could only
  be the case if $\alpha\gtrsim{}2$.
\item Magnetic fields $\sim{}1~\rm{}mG$ would be required to support
  some of the fragments with $\alpha\ll{}2$ against collapse
  (Sec.~\ref{sec:magnetic-fields}). Fields of this strength would be
  marginally consistent with observations of magnetic fields.
\item Competitive accretion (e.g.,
  \citealt{bonnell1997:mass-spectrum}) might play a role in star
  formation (Sec.~\ref{sec:competitive-accretion}). That process
  requires high gas densities and low velocity dispersions, implying
  $\alpha<1$ \citep{krumholz2005:non-competitive}, just as observed in
  many cloud fragments.
\end{itemize}
These conclusions hinge on the assumption that the observed virial
parameters are relatively free of biases, and that theory correctly
predicts critical virial parameters. Both assumptions are unlikely to
be wrong by factors of order 10, as needed to explain all observed
virial parameters without recourse to models appropriate for
$\alpha\ll{}2$ (Sec.~\ref{sec:vp_correct-relevant}).

The scatter in the virial parameter seen in Fig.~\ref{fig:vp-observed}
also implies that the virial mass is a very crude tool with which to
assess cloud masses (Sec.~\ref{sec:virial-mass}). To obtain good mass
estimates from a virial analysis, $\alpha{}\approx{}1$ must hold. As
we see in Fig.~\ref{fig:vp-observed}, this is not the case for a large
number of cloud fragments. For an individual cloud or cloud fragment,
$M$ and $M_{\rm{}vir}$ can differ by a factor of 10 or more.

\acknowledgements{We are grateful to a very constructive and helpful
  anonymous referee with attention to detail. The reports
  significantly helped to improve the readability of the paper, and
  also helped to streamline some of the quantitative arguments
  presented in this work. We thank Christopher F.\ McKee, Jonathan
  Tan, and Matthew Bate for reading and commenting on our manuscript
  in advance of publication; Marion Wienen for kindly providing data
  before publication; and Kostas Tassis for enlightening discussions
  on magnetic fields in SF regions. We like to thank further
  colleagues who helped to develop this paper over several stages:
  Fumitaka Nakamura, Norman Murray, Fred C.\ Adams, and Tom
  Megeath. JK acknowledges support from the NASA Astrophysics Data
  Analysis program. This paper benefited from invitations of TP and JK
  to: the 2012 ASTROWIN workshop at the University of Florida at
  Gainesville, organized by Jonathan Tan; the 2012 Star Formation
  Winter School held at the National Astronomical Observatory of Japan
  (NAOJ) in Tokyo, organized by Fumitaka Nakamura; and Caltech's Red
  Door Caffee. TP and JK dedicate this paper to the memory of KSS.}

\bibliographystyle{apj}
\bibliography{/Users/jens/texinputs/mendeley/library}

\begin{thebibliography}{81}
\expandafter\ifx\csname natexlab\endcsname\relax\def\natexlab#1{#1}\fi

\bibitem[{Adams(2010)}]{adams2010:review}
Adams, F.~C. 2010, \araa, 48, 47

\bibitem[{Alves {et~al.}(2007)Alves, Lombardi, \& Lada}]{alves2007:imf}
Alves, J., Lombardi, M., \& Lada, C. 2007, \aap, 462, L17

\bibitem[{Ballesteros-Paredes(2006)}]{ballesteros-paredes2006:six-myths}
Ballesteros-Paredes, J. 2006, \mnras, 372, 443

\bibitem[{Barranco \& Goodman(1998)}]{barranco1998:coherent_cores}
Barranco, J., \& Goodman, A. 1998, \apj, 504, 207

\bibitem[{Bate {et~al.}(2002)Bate, Bonnell, \& Bromm}]{bate2002:brown-dwarfs}
Bate, M., Bonnell, I., \& Bromm, V. 2002, \mnras, 332, L65

\bibitem[{Bate \& Bonnell(2005)}]{bate2005:imf}
Bate, M.~R., \& Bonnell, I.~A. 2005, \mnras, 356, 1201

\bibitem[{Battersby {et~al.}(2011)Battersby, Bally, Ginsburg, Bernard, Brunt,
  Fuller, Martin, Molinari, Mottram, Peretto, Testi, \&
  Thompson}]{battersby2011:cluster-precursors}
Battersby, C., Bally, J., Ginsburg, A., et al. 2011, \aap, 535, A128

\bibitem[{Beltran {et~al.}(2004)Beltran, Cesaroni, Neri, Codella, Furuya,
  Testi, \& Olmi}]{beltran2004:rotating-disks}
Beltran, M.~T., Cesaroni, R., Neri, R., et al. 2004, \apj, 601, L187

\bibitem[{Bertoldi \& McKee(1992)}]{bertoldi1992:pr_conf_cores}
Bertoldi, F., \& McKee, C.~F. 1992, \apj, 395, 140

\bibitem[{Beuther {et~al.}(2002)Beuther, Schilke, Menten, Motte, Sridharan, \&
  Wyrowski}]{beuther2002:hmpo-densities}
Beuther, H., Schilke, P., Menten, K.~M., et al. 2002, \apj, 566, 945

\bibitem[{Beuther {et~al.}(2005)Beuther, Sridharan, \&
  Saito}]{beuther2005:hmsf-onset}
Beuther, H., Sridharan, T.~K., \& Saito, M. 2005, \apj, 634, L185

\bibitem[{Bianchi {et~al.}(2003)Bianchi, Gon\c{c}alves, Albrecht, Caselli,
  Chini, Galli, \& Walmsley}]{bianchi2003:dust-emissivity}
Bianchi, S., Gon\c{c}alves, J., Albrecht, M., et al. 2003, \aap, 399, L43

\bibitem[{Bonnell \& Bate(2006)}]{bonnell2006:competitive-accretion}
Bonnell, I.~A., \& Bate, M.~R. 2006, \mnras, 370, 488

\bibitem[{Bonnell {et~al.}(1997)Bonnell, Bate, Clarke, \&
  Pringle}]{bonnell1997:mass-spectrum}
Bonnell, I.~A., Bate, M.~R., Clarke, C.~J., \& Pringle, J.~E. 1997, \mnras,
  285, 201

\bibitem[{Bonnell {et~al.}(2001{\natexlab{a}})Bonnell, Bate, Clarke, \&
  Pringle}]{bonnell2001:competitive-accretion}
---. 2001{\natexlab{a}}, \mnras, 323, 785

\bibitem[{Bonnell {et~al.}(2001{\natexlab{b}})Bonnell, Clarke, Bate, \&
  Pringle}]{bonnell2001:accretion-imf}
Bonnell, I.~A., Clarke, C.~J., Bate, M.~R., \& Pringle, J.~E.
  2001{\natexlab{b}}, \mnras, 324, 573

\bibitem[{Bonnor(1956)}]{bonnor1956:be-spheres}
Bonnor, W. 1956, \mnras, 116, 351

\bibitem[{Bontemps {et~al.}(2010)Bontemps, Motte, Csengeri, \&
  Schneider}]{bontemps2010:cygnus-x}
Bontemps, S., Motte, F., Csengeri, T., \& Schneider, N. 2010, \aap, 524, A18

\bibitem[{Carey {et~al.}(1998)Carey, Clark, Egan, Price, Shipman, \&
  Kuchar}]{carey1998:irdc-properties}
Carey, S.~J., Clark, F.~O., Egan, M.~P., et al. 1998, \apj, 508, 721

\bibitem[{Caselli {et~al.}(2002)Caselli, Benson, Myers, \&
  Tafalla}]{caselli2002:n2h+}
Caselli, P., Benson, P., Myers, P., \& Tafalla, M. 2002, \apj, 572, 238

\bibitem[{Crutcher(2012)}]{crutcher2012:review}
Crutcher, R.~M. 2012, \araa, 50, 29

\bibitem[{Csengeri {et~al.}(2011)Csengeri, Bontemps, Schneider, Motte, \&
  Dib}]{csengeri2011:cygnus-x}
Csengeri, T., Bontemps, S., Schneider, N., Motte, F., \& Dib, S. 2011, \aap,
  527, A135

\bibitem[{Dobbs {et~al.}(2011)Dobbs, Burkert, \&
  Pringle}]{dobbs2011:unbound-clouds}
Dobbs, C.~L., Burkert, A., \& Pringle, J.~E. 2011, \mnras, 413, 2935

\bibitem[{Ebert(1955)}]{ebert1955:be-spheres}
Ebert, R. 1955, Zeitschrift f\"ur Astrophysik, 37, 217

\bibitem[{Egan {et~al.}(1998)Egan, Shipman, Price, Carey, Clark, \&
  Cohen}]{egan1998:irdcs}
Egan, M.~P., Shipman, R.~F., Price, S.~D., et al. 1998, \apjl, 494, L199

\bibitem[{Enoch {et~al.}(2006)Enoch, Young, Glenn, {Evans II}, Golwala,
  Sargent, Harvey, Aguirre, Goldin, Haig, Huard, Lange, Laurent, Maloney,
  Mauskopf, Rossinot, \& Sayers}]{enoch2006:perseus}
Enoch, M., Young, K., Glenn, J., et al. 2006, \apj, 638, 293

\bibitem[{Evans {et~al.}(2009)Evans, Dunham, J\o~rgensen, Enoch, Mer\'{\i}n,
  van Dishoeck, Alcal\'{a}, Myers, Stapelfeldt, Huard, Allen, Harvey, van
  Kempen, Blake, Koerner, Mundy, Padgett, \& Sargent}]{evans2009:c2d-summary}
Evans, N.~J., Dunham, M.~M., J\o~rgensen, J.~K., et al. 2009, \apjs,
  181, 321

\bibitem[{Fontani {et~al.}(2005)Fontani, Beltr�n, Brand, Cesaroni, Testi,
  Molinari, \& Walmsley}]{fontani2005:southern-hmpo-candidates}
Fontani, F., Beltran, M.~T., Brand, J., et al. 2005, \aap, 432, 921

\bibitem[{Foster {et~al.}(2009)Foster, Rosolowsky, Kauffmann, Pineda, Borkin,
  Caselli, Myers, \& Goodman}]{foster2008:nh3-analysis}
Foster, J.~B., Rosolowsky, E.~W., Kauffmann, J., et al. 2009, \apj, 696, 298

\bibitem[{Frerking {et~al.}(1982)Frerking, Langer, \&
  Wilson}]{frerking1982:abundance}
Frerking, M.~A., Langer, W.~D., \& Wilson, R.~W. 1982, \apj, 262, 590

\bibitem[{Goldsmith {et~al.}(2008)Goldsmith, Heyer, Narayanan, Snell, Li, \&
  Brunt}]{goldsmith2008:taurus-fcrao}
Goldsmith, P., Heyer, M., Narayanan, G., et al.. 2008,
  \apj, 680, 428

\bibitem[{Goodman {et~al.}(1998)Goodman, Barranco, Wilner, \&
  Heyer}]{goodman1998:coherent_cores}
Goodman, A., Barranco, J., Wilner, D., \& Heyer, M. 1998, \apj, 504, 223

\bibitem[{Goodman {et~al.}(1993)Goodman, Benson, Fuller, \&
  Myers}]{goodman1993:velocity-gradients}
Goodman, A.~A., Benson, P.~J., Fuller, G.~A., \& Myers, P.~C. 1993, \apj, 406,
  528

\bibitem[{Hennebelle {et~al.}(2011)Hennebelle, Commer\c{c}on, Joos, Klessen,
  Krumholz, Tan, \& Teyssier}]{hennebelle2011:collapse-outflows-fragmentation}
Hennebelle, P., Commer\c{c}on, B., Joos, M., et al. 2011, \aap, 528, A72

\bibitem[{Heyer {et~al.}(2009)Heyer, Krawczyk, Duval, \&
  Jackson}]{heyer2009:scaling-relations}
Heyer, M., Krawczyk, C., Duval, J., \& Jackson, J. 2009, \apj, 699, 1092

\bibitem[{Heyer {et~al.}(2001)Heyer, Carpenter, \&
  Snell}]{heyer2001:equilibrium}
Heyer, M.~H., Carpenter, J.~M., \& Snell, R.~L. 2001, \apj, 551, 852

\bibitem[{Hill {et~al.}(2011)Hill, Motte, Didelon, Bontemps, Minier, Hennemann,
  Schneider, Andr\'{e}, Men‘shchikov, Anderson, Arzoumanian, Bernard,
  di~Francesco, Elia, Giannini, Griffin, K\"{o}nyves, Kirk, Marston, Martin,
  Molinari, {Nguyen Luong}, Peretto, Pezzuto, Roussel, Sauvage, Sousbie, Testi,
  Ward-Thompson, White, Wilson, \& Zavagno}]{hill2011:vela-c}
Hill, T., Motte, F., Didelon, P., et al. 2011, \aap, 533, A94

\bibitem[{Kauffmann {et~al.}(2008)Kauffmann, Bertoldi, Bourke, Evans, \&
  Lee}]{kauffmann2008:mambo-spitzer}
Kauffmann, J., Bertoldi, F., Bourke, T.~L., Evans, N.~J., \& Lee, C.~W. 2008,
  \aap, 487, 993

\bibitem[{Kauffmann \& Pillai(2010)}]{jenthu2010:irdcs}
Kauffmann, J., \& Pillai, T. 2010, \apj, 723, L7

\bibitem[{Kauffmann {et~al.}(2010{\natexlab{a}})Kauffmann, Pillai, Shetty,
  Myers, \& Goodman}]{kauffmann2010:mass-size-i}
Kauffmann, J., Pillai, T., Shetty, R., Myers, P.~C., \& Goodman, A.~A.
  2010{\natexlab{a}}, \apj, 712, 1137

\bibitem[{Kauffmann {et~al.}(2010{\natexlab{b}})Kauffmann, Pillai, Shetty,
  Myers, \& Goodman}]{kauffmann2010:mass-size-ii}
---. 2010{\natexlab{b}}, \apj, 716, 433

\bibitem[{Kirk {et~al.}(2010)Kirk, Pineda, Johnstone, \&
  Goodman}]{kirk2010:dense-core-dynamics-ii}
Kirk, H., Pineda, J.~E., Johnstone, D., \& Goodman, A. 2010, \apj, 723, 457

\bibitem[{Kr\"{u}gel \& Siebenmorgen(1994)}]{kruegel1994:opacities}
Kr\"{u}gel, E., \& Siebenmorgen, R. 1994, \aap, 288, 929

\bibitem[{Krumholz {et~al.}(2005)Krumholz, McKee, \&
  Klein}]{krumholz2005:non-competitive}
Krumholz, M.~R., McKee, C.~F., \& Klein, R.~I. 2005, \nat, 438, 332

\bibitem[{Lada {et~al.}(2008)Lada, Muench, Rathborne, Alves, \&
  Lombardi}]{lada2008:pipe-nature}
Lada, C., Muench, A., Rathborne, J., Alves, J., \& Lombardi, M. 2008, \apj,
  672, 410

\bibitem[{Larson(1981)}]{larson1981:linewidth_size}
Larson, R. 1981, \mnras, 194, 809

\bibitem[{Li {et~al.}(2013)Li, Kauffmann, Zhang, \& Chen}]{li2012:orion}
Li, D., Kauffmann, J., Zhang, Q., \& Chen, W. 2013, \apj, 768, L5

\bibitem[{{Mac Low} \& Klessen(2004)}]{maclow2004:review}
{Mac Low}, M.-M., \& Klessen, R. 2004, Reviews of Modern Physics, 76, 125

\bibitem[{Maloney(1990)}]{maloney1990:virial-equilibrium}
Maloney, P. 1990, \apj, 348, L9

\bibitem[{McKee \& Holliman(1999)}]{mckee1999:mpp}
McKee, C., \& Holliman, J. 1999, \apj, 522, 313

\bibitem[{McKee \& Ostriker(2007)}]{mckee2007:review}
McKee, C., \& Ostriker, E. 2007, \araa, 45, 565

\bibitem[{McKee \& Tan(2002)}]{mckee2002:massive-sf}
McKee, C.~F., \& Tan, J.~C. 2002, \nat, 416, 59

\bibitem[{McKee \& Tan(2003)}]{mckee2003:massive-sf}
---. 2003, \apj, 585, 850

\bibitem[{Milam {et~al.}(2005)Milam, Savage, Brewster, Ziurys, \&
  Wyckoff}]{milam2005:co-isotope-abundance}
Milam, S.~N., Savage, C., Brewster, M.~A., Ziurys, L.~M., \& Wyckoff, S. 2005,
  \apj, 634, 1126

\bibitem[{Molinari {et~al.}(2000)Molinari, Brand, Cesaroni, \&
  Palla}]{molinari2000:hii-precursors}
Molinari, S., Brand, J., Cesaroni, R., \& Palla, F. 2000, \aap

\bibitem[{Motte {et~al.}(1998)Motte, Andre, \& Neri}]{motte1998:ophiuchus}
Motte, F., Andre, P., \& Neri, R. 1998, \aap, 336, 150

\bibitem[{Myers \& Goodman(1988)}]{myers1988:mhd-equilibrium}
Myers, P.~C., \& Goodman, A.~A. 1988, \apj, 326, L27

\bibitem[{Nutter \& Ward-Thompson(2007)}]{nutter2007:orion}
Nutter, D., \& Ward-Thompson, D. 2007, \mnras, 374, 1413

\bibitem[{Ossenkopf \& Henning(1994)}]{ossenkopf1994:opacities}
Ossenkopf, V., \& Henning, T. 1994, \aap, 291, 943

\bibitem[{Pillai {et~al.}(2011)Pillai, Kauffmann, Wyrowski, Hatchell, Gibb, \&
  Thompson}]{pillai2011:initial-conditions}
Pillai, T., Kauffmann, J., Wyrowski, F., et al. 2011, \aap, 530, A118

\bibitem[{Pillai {et~al.}(2006)Pillai, Wyrowski, Menten, \&
  Kr\"{u}gel}]{pillai2006:g11}
Pillai, T., Wyrowski, F., Menten, K., \& Kr\"{u}gel, E. 2006, \aap, 447, 929

\bibitem[{Plume {et~al.}(1997)Plume, Jaffe, {Evans II}, Martin‐Pintado, \&
  Gomez‐Gonzalez}]{plume1997:water-maser-cores}
Plume, R., Jaffe, D.~T., {Evans II}, N.~J., Martin‐Pintado, J., \&
  Gomez‐Gonzalez, J. 1997, \apj, 476, 730

\bibitem[{Ragan {et~al.}(2009)Ragan, Bergin, \&
  Gutermuth}]{ragan2009:irdc-extinction}
Ragan, S., Bergin, E., \& Gutermuth, R. 2009, \apj, 698, 324

\bibitem[{Ragan {et~al.}(2012)Ragan, Heitsch, Bergin, \&
  Wilner}]{ragan2012:irdcs-vla}
Ragan, S.~E., Heitsch, F., Bergin, E.~A., \& Wilner, D. 2012, \apj, 746, 174

\bibitem[{Rathborne {et~al.}(2008)Rathborne, Lada, Muench, Alves, \&
  Lombardi}]{rathborne2008:pipe-nh3}
Rathborne, J., Lada, C., Muench, A., Alves, J., \& Lombardi, M. 2008, \apjs,
  174, 396

\bibitem[{Rathborne {et~al.}(2007)Rathborne, Simon, \&
  Jackson}]{rathborne2007:irdc-msf}
Rathborne, J.~M., Simon, R., \& Jackson, J.~M. 2007, \apj, 662, 1082

\bibitem[{Reid {et~al.}(2009)Reid, Menten, Zheng, Brunthaler, Moscadelli, Xu,
  Zhang, Sato, Honma, Hirota, Hachisuka, Choi, Moellenbrock, \&
  Bartkiewicz}]{reid2009:parallaxes}
Reid, M.~J., Menten, K.~M., Zheng, X.~W., et al. 2009, \apj, 700, 137

\bibitem[{Roman-Duval {et~al.}(2010)Roman-Duval, Jackson, Heyer, Rathborne, \&
  Simon}]{roman-duval2010:grs-clouds}
Roman-Duval, J., Jackson, J.~M., Heyer, M., Rathborne, J., \& Simon, R. 2010,
  \apj, 723, 492

\bibitem[{Rosolowsky {et~al.}(2008)Rosolowsky, Pineda, Foster, Borkin,
  Kauffmann, Caselli, Myers, \& Goodman}]{rosolowsky2007:perseus_nh3}
Rosolowsky, E., Pineda, J., Foster, J., et al. 2008, \apjs, 175, 509

\bibitem[{Scoville {et~al.}(1987)Scoville, Yun, Sanders, Clemens, \&
  Waller}]{scoville1987:inner-galaxy-clouds}
Scoville, N.~Z., Yun, M.~S., Sanders, D.~B., Clemens, D.~P., \& Waller, W.~H.
  1987, \apjs, 63, 821

\bibitem[{Shu(1977)}]{shu1977:self-sim_collapse}
Shu, F. 1977, \apj, 214, 488

\bibitem[{Solomon {et~al.}(1987)Solomon, Rivolo, Barrett, \&
  Yahil}]{solomon1987:scaling-relations}
Solomon, P., Rivolo, A., Barrett, J., \& Yahil, A. 1987, \apj, 319, 730

\bibitem[{Sridharan {et~al.}(2005)Sridharan, Beuther, Saito, Wyrowski, \&
  Schilke}]{sridharan2005:high-mass-starless-cores}
Sridharan, T.~K., Beuther, H., Saito, M., Wyrowski, F., \& Schilke, P. 2005,
  \apj, 634, L57

\bibitem[{Swift(2009)}]{swift2009:starless-irdc}
Swift, J.~J. 2009, \apj, 705, 1456

\bibitem[{Tan {et~al.}(2013)Tan, Kong, Butler, Caselli, \&
  Fontani}]{tan2013:hmsc}
Tan, J.~C., Kong, S., Butler, M.~J., Caselli, P., \& Fontani, F. 2013, eprint
  arXiv:1303.4343

\bibitem[{Tomisaka {et~al.}(1988)Tomisaka, Ikeuchi, \& Nakamura}]{tin_ii}
Tomisaka, K., Ikeuchi, S., \& Nakamura, T. 1988, \apj, 335, 239

\bibitem[{Wang {et~al.}(2006)Wang, Zhang, Rathborne, Jackson, \&
  Wu}]{wang2006:irdc-masers}
Wang, Y., Zhang, Q., Rathborne, J.~M., Jackson, J., \& Wu, Y. 2006, \apj, 651,
  L125

\bibitem[{Wienen {et~al.}(2012)Wienen, Wyrowski, Schuller, Menten, Walmsley,
  Bronfman, \& Motte}]{wienen2012:ammonia}
Wienen, M., Wyrowski, F., Schuller, F., et al. 2012, \aap, 544, A146

\bibitem[{Williams {et~al.}(2000)Williams, Blitz, \&
  McKee}]{williams2000:pp_iv}
Williams, J., Blitz, L., \& McKee, C. 2000, Protostars and Planets IV, 97

\bibitem[{Zhang \& Wang(2011)}]{zhang2011:tale-two-cores}
Zhang, Q., \& Wang, K. 2011, \apj, 733, 26

\bibitem[{Zinnecker \& Yorke(2007)}]{zinnecker2007:msf-review}
Zinnecker, H., \& Yorke, H.~W. 2007, \araa, 45, 481

\end{thebibliography}

\appendix

\section{The Virial Parameter: Non--Isothermal Models and Complex
  Density Distributions}
This appendix summarizes advanced aspects of the virial parameter
discussion that had to be skipped in Sec.~\ref{sec:vp-summary}. In
particular, we consider non--magnetized spheres supported by
non--isothermal pressure, as well as deviations from the assumed
density profiles.\label{app:vp-calculations}\medskip

The Bonnor--Ebert model from Sec.~\ref{sec:vp-summary} provides a
helpful reference case for stability considerations. In the original
discussion, the velocity dispersion entering the Bonnor--Ebert mass,
$\sigma_v$, measures the gas temperature. It can also be taken to
represent a---spatially constant---velocity dispersion due to random
non--thermal ``turbulent'' gas motions.

In practice, though, velocity dispersions are usually not spatially
constant within molecular clouds. The velocity dispersion typically
increases with increasing spatial scale (e.g.,
\citealt{goodman1998:coherent_cores}). Given density gradients, this
means the velocity dispersion is anticorrelated with the gas
density. One often parameterizes these trends using a polytropic
equation of state relating pressure and density,
$P\propto{}\varrho^{\gamma_P}$. Provided the pressure comes from
random gas motions, $P(r)=\varrho(r)\,\sigma_v^2(r)$, where a
dependence on radius $r$ is considered. In this case, the velocity
dispersion obeys $\sigma_v(r)\propto{}\varrho(r)^{(\gamma_P-1)/2}$ and
is anticorrelated with the density for polytropic exponents
$\gamma_P<1$. Such polytropes are, for example, considered by
\citet{mckee1999:mpp}. Combination of their Eqs.~(64, 67) shows that
\begin{equation}
M_{\rm{}cr} \le{} 2.47\,\frac{\langle{}\sigma_v^2\rangle{}\,R}{G}\,
\left(\frac{2-\gamma_P}{4-3\gamma_P}\right)^{1/3}
\end{equation}
is the critical mass for pressure--confined hydrostatic equilibrium
spheres supported by non--thermal pressure characterized by
$\gamma_P<1$, where $\langle{}\sigma_v^2\rangle$ represents a
mass--weighted average. Substitution in
Eq.~(\ref{eq:virial-parameter-definition}) results in critical virial
parameters $\alpha_{\rm{}cr}\ge{}2.02$ for polytropic exponents in the
range $0\le{}\gamma_P<1$. This implies
$\alpha_{B=0}\gtrsim{}\alpha_{\rm{}BE}\approx{}2$ and
$M_{B=0}\lesssim{}M_{\rm{}BE}$, as already mentioned in
Sec.~\ref{sec:vp-summary}.\medskip

Cloud fragments are, of course, not necessarily well approximated as
spheres. Also, internal density gradients may differ from those
prevailing in the aforementioned idealized models. The impact of these
factors is absorbed into the parameter $a$. For spheroidal mass
distributions, \citetalias{bertoldi1992:pr_conf_cores} demonstrate
that the impact of deviations from a spherical state, $a_{\vartheta}$,
can be separated from the influence of density gradients,
$a_{\varrho}$, and that it is possible to write
$a=a_{\vartheta}\cdot{}a_{\varrho}$.

To parametrize our problem, we follow
\citetalias{bertoldi1992:pr_conf_cores} in their assumption that the
considered gaseous body is a triaxial ellipsoid: the extent along
two semi--axes is assumed to be identical, and gives the equatorial
radius, $R_{\rm{}eq}$. The extent along the third semi--axis is $Z$:
the body is oblate (``pancake--shaped'') for $Z<R_{\rm{}eq}$, a sphere
in the case $Z=R_{\rm{}eq}$, and prolate (``cigar--shaped'') when
$Z>R_{\rm{}eq}$. For a given projection on the plane of the sky, with
projected semi--axes $R_{\rm{}min}$ and $R_{\rm{}max}$, the observed
radius is $R_{\rm{}obs}=(R_{\rm{}min}\cdot{}R_{\rm{}max})^{1/2}$. We
only summarize the results of calculations established by
\citetalias{bertoldi1992:pr_conf_cores}. We refer to that publication
for details.

First, consider the ratio between the true gravitational potential
energy $E_{\rm{}pot}$, calculated from the three--dimensional shape of
the body, and the ``observed'' energy $E_{\rm{}pot,obs}$, calculated
from $R_{\rm{}obs}$ and the mass. Assuming a random projection onto
the sky, and $Z/R_{\rm{}eq}<10$, we find that
$|E_{\rm{}pot}|/|E_{\rm{}pot,obs}|=1.0\pm{}0.5$ for 80\% of all random
viewing directions. In other words, it is
possible to estimate the potential energy of a cloud fragment with a
reasonable degree of reliability.

The same set of calculations also permits to estimate how much
$R_{\rm{}obs}$ differs between different statistical realizations of
the projection. For a given axis ratio $Z/R_{\rm{}eq}$, we use the
average projected radius $\langle{}R_{\rm{}obs}\rangle$ as a
reference.  Again, for 90\% of the cases,
$R_{\rm{}obs}/\langle{}R_{\rm{}obs}\rangle{}=1.0\pm{}0.5$ in the range
$Z/R_{\rm{}eq}<10$. This excludes, however, very small ratios
$Z/R_{\rm{}eq}\lesssim{}0.2$, where deviations are significantly
larger.

Second, consider the impact of density gradients. Assume that the body
consists of ellipsoidal density shells, so that the density
gradient is $\varrho\propto{}r^{-k}$ along any radius vector. In that
case, $a_{\varrho}=(1-k/3)/(1-2k/5)$, following \citetalias{bertoldi1992:pr_conf_cores}. This gives
$a_{\varrho}\to{}1$ for $k\to{}0$, and $a_{\varrho}\to{}\infty$ for
$k\to{}5/2$. A fiducial density gradient $k=2$ gives
$a_{\varrho}=5/3$. For $0<k<2.3$, values $a_{\varrho}=2\pm{}1$ hold.

\section{The Virial Parameter Slope as a Consequence from Mass--Size
  and Linewidth--Size Laws}
It is straightforward to show that
mass--size and linewidth--size laws imply the observed virial
parameter slope. Consider the definition of the virial parameter,
given in Eq.\ (\ref{eq:virial-parameter-definition}). Logarithmic
differentiation yields
\begin{eqnarray}
\d \log(\alpha) & = & 2 \d \log(\sigma_v) + \d \log(R) - \d \log(M)\\
 & = & 2 \frac{\d \log(\sigma_v)}{\d \log(R)} \d \log(R) + \d \log(R)
       - \d \log(M)\\
 & = & \left( 2 \frac{\d \log(\sigma_v)}{\d \log(R)} + 1 -
       \frac{\d \log(M)}{\d \log(R)} \right)
       \frac{\d \log(R)}{\d \log(M)} \d \log(M) \, .
\end{eqnarray}
Rearrangement gives
\begin{equation}
\frac{\d \log(\alpha)}{\d \log(M)} =
       \frac{
         2 \frac{\d \log(\sigma_v)}{\d \log(R)} + 1 -
         \frac{\d \log(M)}{\d \log(R)}
       }{
         \d \log(M) / \d \log(R)
       } \, .
\end{equation}
Obviously, the virial parameter slope, $\d{}\log(\alpha)/\d{}\log(M)$,
is a direct result of the relevant mass--size
slope, $\d{}\log(M)/\d{}\log(R)$, and the linewidth--size relation,
$\d{}\log(M)/\d{}\log(R)$. Section~\ref{sec:vp-slope} explores this
result.\label{app:vp-slope}

\section{Implications of Low Virial Parameters}
This appendix includes several calculations relevant to interpretation
of the implications of low virial
parameters. Section~\ref{sec:sf-physics} applies the calculations to
the observations.\label{app:implications}

\subsection{Low Virial Parameters are characteristic of HMSF}
In the limit $\sigma_v=\sigma_{v,\rm{}nt}$, substitution of the
mass--size threshold for HMSF (Eq.~\ref{eq:threshold-hmsf}) and the
linewidth--size relation (Eq.~\ref{eq:linewidh-size}) in
Eq.~(\ref{eq:virial-parameter}) gives
\begin{equation}
\alpha = 1.3 \,
\left( \frac{M}{M_{\rm{}HMSF}(R)} \right)^{-1} \,
\left( \frac{\sigma_{v,0}}{0.8~\rm{}km\,s^{-1}} \right)^{2} \,
\left( \frac{R}{0.1~\rm{}pc} \right)^{0.31} \, .
\label{eq:low-alpha-hmsf}
\end{equation}
Since $\sigma_{v,0}\approx{}0.8~\rm{}km\,s^{-1}$, virial parameters
$\alpha<1$ typically hold for HMSF fragments where
$M>M_{\rm{}HMSF}(R)$. Likewise, small values of $\alpha$ suggest large
values of $M/M_{\rm{}HMSF}$. The observed small virial parameters are
therefore most relevant for the formation of high--mass stars.

\subsection{Short Lifetimes of High--Mass Starless
  Cores\label{app-sec:absence-starless-cores}}
The free--fall timescale is given by
\begin{equation}
\tau_{\rm{}ff} =
  \left( \frac{3 \pi}{32 \, G \, \langle{}\varrho\rangle} \right)^{1/2}
  = 1.7\times{}10^5 ~ {\rm{}yr}
  \left( \frac{M}{10 \, M_{\sun}} \right)^{-1/2}
  \left( \frac{R}{0.1 ~ \rm{} pc} \right)^{3/2} \, ,
\label{eq:tau-ff}
\end{equation}
where $\langle{}\varrho\rangle{}=M/(4/3\,\pi\,R^3)$ is the mean
density of a sphere. In HMSF regions, mass and size are related by the
approximate threshold for high--mass star formation,
Eq.~(\ref{eq:threshold-hmsf}). When we substitute this relation, we
find that
\begin{equation}
\tau_{\rm{}ff} < 5.5\times{}10^4~{\rm{}yr}
  \left( \frac{M}{10 \, M_{\sun}} \right)^{0.63}
\end{equation}
holds for cloud fragments $M>M_{\rm{}HMSF}$ deemed able to form
high--mass stars.

\subsection{Fragments with Small Virial Parameters are not
  Collapsing\label{app-sec:no-collapse}}
Consider a fragment with $\alpha<\alpha_{\rm{}cr}$ with an initial
energy content $E_{\rm{}kin,0}$, $E_{\rm{}pot,0}$. This fragment will
begin to contract because it is supercritical. Unless some process
drains energy from the fragment, conservation of energy requires that
the potential energy released by contraction will lead to an increase
in kinetic energy, i.e.,
$\Delta{}E_{\rm{}kin}=\left|{}\Delta{}E_{\rm{}pot}\right|{}$. As a
consequence, the virial parameter in the contracting cloud becomes
\begin{equation}
\alpha = a \,
  \frac{
        2 \, (E_{\rm{}kin, 0} + \Delta E_{\rm{}kin})
       }{
        \left| E_{\rm{}pot, 0} + \Delta E_{\rm{}pot} \right|
       }
\label{eq:alpha-contraction}
\end{equation}
In the limit of contraction to radii much smaller than the initial
radius where contraction started,
$|E_{\rm{}pot,0}|\ll{}|\Delta{}E_{\rm{}pot}|$
and
$E_{\rm{}kin,0}\ll{}\Delta{}E_{\rm{}kin}=\left|{}\Delta{}E_{\rm{}pot}\right|{}$.
Thus,
\begin{equation}
\alpha \to 2 \, a
\end{equation}
follows for cloud fragments having undergone significant
contraction. Typically, $a=2\pm{}1$
(Appendix~\ref{app:vp-calculations}), roughly implying
$\alpha{}\to{}4\pm{}2$ during collapse.

In fact, virial parameters $\ll{}\alpha_{\rm{}cr}$ are not found at
any stage during collapse. Equation~\ref{eq:alpha-contraction} implies
that $\alpha$ monotonously changes from its initial value,
$\alpha_0=2a\,{}E_{\rm{}kin, 0}/|E_{\rm{}pot, 0}|$, to $2a$. Since
collapse is initiated when the cloud fragment reaches the critical
state, it is plausible to assume $\alpha_0=\alpha_{\rm{}cr}$. This
means that, during collapse, the virial parameter is between
$\alpha_{\rm{}cr}$ and $2a$.\medskip

The argument above only holds if energy is conserved. This condition
can only be violated if the timescale for energy losses,
$\tau_{\rm{}loss}$, is shorter than the one for the collapse occurring
in approximate free--fall fashion,
$\tau_{\rm{}loss}<\tau_{\rm{}ff}$. Energy could, for example, be
drained in the form of radiation from hot gas or in the form of
magnetohydrodynamic waves. However, it seems unlikely that
the condition $\tau_{\rm{}loss}<\tau_{\rm{}ff}$ can be fulfilled.

Consider radiation from warm gas. The crossing time,
$\tau_{\rm{}cross}\equiv{}R/\sigma_v$, is the characteristic timescale
for conversion of kinetic energy to heat
\citep{maclow2004:review}. To have significant
energy loss via radiation, we need to have
$\tau_{\rm{}cross}<\tau_{\rm{}ff}$. Now, combining the definition of
the crossing time with the free--fall timescale (Eq.~\ref{eq:tau-ff}),
we can rewrite the virial parameter (Eq.~\ref{eq:virial-parameter-definition})
as
\begin{equation}
\alpha = \frac{45}{32} \,
  \left( \frac{\tau_{\rm{}ff}}{\tau_{\rm{}cross}} \right)^2 \, .
\label{eq:virial-parameter-ff}
\end{equation}
In this section, we consider supercritical cloud fragments
characterized by $\alpha{}\ll{}2$. For this situation,
Eq.~(\ref{eq:virial-parameter-ff}) implies
$\tau_{\rm{}cross}\gg{}\tau_{\rm{}ff}$. This is in conflict with the
condition for significant energy losses from heating radiation during
collapse, $\tau_{\rm{}cross}\lesssim{}\tau_{\rm{}ff}$. It follows that
heating processes cannot help to radiate energy away: this would
require heating in fast gas motions, which however also support the
cloud fragment.

Alternatively, we can consider energy to be radiated away via
magnetohydrodynamic (MHD) waves traveling at the Alfv\'en velocity. The
latter is $v_{\rm{}A}=B/(4\pi\langle{}\varrho\rangle)^{1/2}$, where we
assume that the field is frozen into the mass reservoir at mean
density $\langle{}\varrho\rangle$. To have significant energy
transport via MHD waves during collapse, the Alfv\'enic crossing time
$\tau_{\rm{}A}=R/v_{\rm{}A}$ must be shorter than the free--fall
timescale, i.e., $\tau_{\rm{}A}<\tau_{\rm{}ff}$. Now, the critical
virial parameter for strongly magnetized clouds can be approximated as
\begin{equation}
\alpha_B \to 11.75 \,
  \left( \frac{\tau_{\rm{}A}}{\tau_{\rm{}cross}} \right)^2 \, .
\label{eq:virial-parameter-iv}
\end{equation}
This approximation follows from
$\alpha=\alpha_{\rm{}BE}\cdot{}(M_{\rm{}BE}/M)$
(Sec.~\ref{sec:vp-summary}) in the limit that the critical mass
becomes equal to the magnetic flux mass, $M_B\to{}M_{\Phi}$. This is
expected in the case considered here since fast and relevant energy
flows in MHD waves, i.e.\ $v_{\rm{}A}\gg{}\sigma_v$, only become
relevant for strong magnetization. Then, via substitution of
Eqs.~(\ref{eq:virial-parameter-ff}, \ref{eq:virial-parameter-iv}),
the condition $\alpha\ll{}\alpha_B$ results in
$\tau_{\rm{}ff}\ll{}2.88\,\tau_{\rm{}A}$. Combined with the condition
for significant energy flow during collapse,
$\tau_{\rm{}A}<\tau_{\rm{}ff}$, we obtain
$0.35\,\tau_{\rm{}ff}\ll{}\tau_{\rm{}A}<\tau_{\rm{}ff}$. These
conditions are marginally fulfilled when
$\tau_{\rm{}A}\approx{}\tau_{\rm{}ff}$. While significant energy flows
are possible, highly efficient flows with
$\tau_{\rm{}A}\ll{}\tau_{\rm{}ff}$ are thus ruled out by these
constraints.

\subsection{Evidence for Significant Magnetic
  Fields?\label{app-sec:magnetic-fields}}
\citet{crutcher2012:review} suggests an approximate density--dependent
upper limit
$B\lesssim{}150~{\rm{}\mu{}G}\cdot{}(n_{\rm{}H_2}/10^4~{\rm{}cm^{-3}})^{0.65}$,
where $n_{\rm{}H_2}$ is the $\rm{}H_2$ particle density. We obtain
\begin{equation}
B \lesssim{} 336~{\rm{}\mu{}G} \,
  \left( \frac{M}{10\,M_{\sun}} \right)^{0.65}
  \left( \frac{R}{0.1~\rm{}pc} \right)^{-1.95}
\end{equation}
if we replace the density by its mean value, $M/(4/3\,\pi\,R^3)$.

\end{document}